\documentclass[aps,prb,twocolumn,superscriptaddress,floatfix,longbibliography]{revtex4-2}

\usepackage{amsmath,amssymb} 
\usepackage{physics}
\usepackage{bm} 
\usepackage{graphicx} 
\usepackage{comment} 
\usepackage{textcomp} 

\usepackage{enumitem}
\setlist{noitemsep,leftmargin=*,topsep=0pt,parsep=0pt}

\usepackage{xcolor} 
\definecolor{lightgray}{gray}{0.6}
\definecolor{medgray}{gray}{0.4}

\usepackage{hyperref}
\hypersetup{
colorlinks=true,
urlcolor= blue,
citecolor=blue,
linkcolor= blue,
}
\def\CR{\nonumber\\}
\newif\ifptitle
\newif\ifpnumber
\newcounter{para}
\newcommand\ptitle[1]{\par\refstepcounter{para}
{\ifpnumber{\noindent\textcolor{lightgray}{\textbf{\thepara}}\indent}\fi}
{\ifptitle{\textbf{[{#1}]}}\fi}}

\newcommand{\mytitle}{Transport of Majorana Bound State in the presence of telegraph noise}

\begin{document}

\title{\mytitle}

\author{Dibyajyoti Sahu}
\email[]{dibyajyoti20@iiserb.ac.in}
\author{Suhas Gangadharaiah}

\email[]{suhasg@iiserb.ac.in}
\affiliation{Department of Physics, IISER Bhopal, 462066, Bhopal India}
\date{\today}

\begin{abstract}
Majorana Bound States (MBS) have emerged as promising candidates for robust quantum computing due to their non-Abelian statistics and topological protection. In this study, we focus on the dynamical transport of MBS in the semiconductor-superconductor (SM-SC) heterostructure via the piano key-type of setup, wherein each of the keys of the wire can be tuned from topological to trivial phases. We focus on the transport of MBS under noisy conditions, and evaluate the feasibility for realistic scenarios. The central emphasis of our work lies in using both numerical and analytical techniques to understand the effect of noise in inducing diabatic errors during transport and to establish scaling laws that relate these errors to the drive time. To achieve this, we derive an effective model that captures the scaling behavior in both noise-free and noisy scenarios, providing a unified framework for analyzing the transport dynamics. We investigate the optimal number of keys for both noisy and noiseless scenarios. Additionally, we explore the effects of disorder on transport dynamics, highlighting its impact on error scaling and robustness. 
\end{abstract}

\maketitle
\section{\label{sec:Introduction}Introduction}

\ptitle{Introduction}
Among the myriad approaches to fault-tolerant quantum computation, MBS stands out for its intrinsic topological robustness and non-Abelian statistics. Predicted by Kitaev’s one-dimensional topological superconducting model~\cite{kitaev_unpaired_2001}, these exotic quasiparticles localize at the ends of a 1D chain, forming nonlocal fermionic states that are resilient to local perturbations. Their unique braiding properties make them promising building blocks for fault-tolerant quantum computation~\cite{nayak_non-abelian_2008, alicea_new_2012, sarma_majorana_2015}.
Following Kitaev’s theoretical prediction, various physical platforms have been proposed to host Majoranas~\cite{prada_andreev_2020}. Early candidates included fractional quantum Hall states at a filling of $\nu=5/2$~\cite{moore_nonabelions_1991} and spinless topological superconductors with a $p_x + ip_y$ pairing symmetry, where MBS are found in the cores of superconducting vortices~\cite{rice_sr2ruo4_1995}. More recent proposals involve 3D topological insulator surfaces proximitized with s-wave superconductors~\cite{fu_superconducting_2008} and semiconductor systems with strong spin-orbit coupling, such as quantum wires placed in proximity to an s-wave superconductor and subject to a magnetic field~\cite{sau_generic_2010, oreg_helical_2010, stanescu_proximity_2010,lutchyn_majorana_2010,dmytruk_renormalization_2018}. This last configuration, particularly in InAs and InSb quantum wires, has seen significant experimental and theoretical progress and holds considerable promise for realizing stable MBS~\cite{albrecht_exponential_2016, churchill_superconductor-nanowire_2013, das_zero-bias_2012, mourik_signatures_2012,reeg_zero-energy_2018,dmytruk_pinning_2020,sahu_effect_2023}.
Despite the lack of definitive evidence for MBS, recent progress in experiments provides significant encouragement.
Parallel to experimental advances, significant theoretical efforts have been devoted to developing practical approaches for leveraging MBS in quantum computation. A key focus is on creating braiding mechanisms that exploit the non-Abelian nature of MZMs to perform quantum gate operations. Proposed strategies include manipulating superconducting wire networks~\cite{alicea_non-abelian_2011, sau_controlling_2011,halperin_adiabatic_2012,tutschku_majorana-based_2020}, exploiting Josephson junction arrays~\cite{heck_coulomb-assisted_2012, hyart_flux-controlled_2013,hegde_topological_2020}, quantum dot chain~\cite{boross_braiding-based_2024,malciu_braiding_2018,liu_andreev_2017,liu_fusion_2023,luethi_majorana_2023,luethi_perfect_2024}, and using periodic driving techniques~\cite{bauer_topologically_2019, martin_double_2020,min_dynamical_2022}. These approaches aim to enable the realization of robust quantum gates, laying the groundwork for practical topological qubits.

\ptitle{Importance of study} In all the physical systems proposed to host MBSs, these quasiparticles are always localized at the boundary between topological and trivial regions. Manipulating the boundary provides a way to control and move MBSs, which is crucial for their potential applications in quantum computing. However, in dynamic situations, no matter the method used to transport MBSs, the time scale of this manipulation is critical. Transport performed on time scales comparable to or shorter than the inverse energy gap risks inducing diabatic transitions, leading to quasiparticle excitations from the ground-state subspace to higher-energy states. These transitions reduce fidelity, introduce decoherence~\cite{lai_decoherence_2020}, and undermine the topological protection essential for robust quantum computation. The impact of these errors becomes even more pronounced in the presence of noise and disorder~\cite{boross_dephasing_2022,boross_braiding-based_2024}, which exacerbate the likelihood of excitations and further degrade system performance. Understanding and mitigating these errors is thus essential for improving MBS transport and advancing toward the realization of topological qubits.

\ptitle{Literature Survey} 
Errors originating from such diabatic transitions have been studied extensively in the context of transport~\cite{scheurer_nonadiabatic_2013,karzig_boosting_2013,karzig_optimal_2015,bauer_dynamics_2018,conlon_error_2019,coopmans_protocol_2021,xu_transport_2022} and braiding protocols~\cite{nag_diabatic_2019,mascot_many-body_2023,cheng_nonadiabatic_2011,karzig_shortcuts_2015,knapp_nature_2016,hell_time_2016,rahmani_optimal_2017,sekania_braiding_2017,zhang_effects_2019}. The transport of MBSs and the associated errors have been explored primarily using Kitaev's spinless 1D toy model. A typical approach involves the piano-key setup, where the wire is divided into discrete segments or "keys." The parameters within each segment can be independently tuned, allowing for precise control of the domain wall and enabling MBS transport along the wire. Studies in this framework have analyzed the effect of the number of keys and drive time on errors, identifying configurations that minimize diabatic transitions~\cite{truong_optimizing_2023}. However, extending this analysis to realistic systems, such as semiconductor-superconductor (SM-SC) heterostructures~\cite{bauer_dynamics_2018,wang_transport_2024}, presents new challenges. In the full model, controlling the transition between trivial and topological regimes is significantly more complex, requiring precise manipulation of system parameters. Additionally, the Majorana localization length in SM-SC heterostructures is comparable to the wire length, which can complicate transport and error mitigation. The dependence of errors on drive time in the presence of noise and disorder also remains less examined, especially in practical implementations of these systems.

\ptitle{Our work and findings}
In this work, we investigate the transport of MBS in a SM-SC heterostructure using a piano key setup to analyze error behavior as a function of drive time. The study employs ground-state overlap as the primary metric, which provides a more accurate measure of defect generation compared to single-state overlaps. 
We observe that the diabatic error initially decreases exponentially with increasing drive time but eventually transitions to a power-law decay which is similar to the spinless case. 
We simulate noisy fluctuations by incorporating symmetric dichotomous temporal noise and study the error generation for various drive times.  For longer drive times, noise significantly amplifies defect production as the system spends more time near critical points, where noise can introduce additional excitations, leading to a substantial increase in diabatic error. In contrast, shorter drive times reduce noise effects, resulting in minimal deviations from the noiseless case. This behavior highlights the existence of an optimal drive time where the effect of noise is minimized. Beyond this point, the diabatic error increases with drive time due to enhanced defect production. We also develop an effective model to predict error scaling in both noisy and noiseless scenarios.
Furthermore, we examine how the number of keys affects diabatic error, observing that increasing the number of keys does not always yield improvements. Instead, there exists an optimal number of keys for a given range of drive times. Specifically, in noiseless, slow-driving regimes, a single key achieves high fidelity, while under noisy conditions, additional keys are beneficial, with four keys proving optimal in our setup. Finally, we investigate the effects of disorder and inhomogeneity on error production during the transport of Majorana modes. Disorder and inhomogeneity not only increases the average error but also influences error production in a complex, non-uniform manner, further complicating the dynamics of Majorana transport.

The paper is organized as follows: In Section \ref{sec:Model}, we describe the SM-SC model and introduce the concept of ground-state adiabatic fidelity. Section \ref{sec:Piano_s_key} focuses on the piano key setup, where we analyze single key to study its effect on the spectrum and diabatic error. Subsection \ref{subsec:s_key_noise} investigates the effect of noise on diabatic error, while Subsection \ref{subsec:s_key_effective} develops an effective model for the SM-SC heterostructure under the single-key setup to predict error scaling in noiseless and noisy scenarios. Section \ref{sec:piano_m_key} extends the study to multiple keys, with Subsection \ref{subsec:optimal_key} addressing the optimal key configuration for a range of drive times. Section \ref{sec:disorder} explores the effects of disorder and inhomogeneity on diabatic error. Finally, Section \ref{sec:conclusion} summarizes the findings and presents the conclusions.

\section{\label{sec:Model}Model}
\ptitle{Model} We investigate a semiconductor-superconductor heterostructure designed to support Majorana bound states (MBS). 
The model consists of a semiconductor with spin-orbit coupling placed next to an s-wave superconductor, which induces an effective superconducting gap in the semiconductor via the proximity effect. A magnetic field is applied perpendicular to the semiconductor wire and parallel to the plane of the superconductor, enabling the necessary conditions for MBS formation. The Hamiltonian, which accounts for all these factors, is expressed as follows:
	\begin{eqnarray}
		&& H =-  \sum_{n}\Bigg\{ \Big[ \sum_{\sigma, \sigma'} c_{n+1,\sigma}^{\dagger}(h_n \delta_{\sigma,\sigma'}-i \alpha_n \sigma_{\sigma, \sigma'}^y)c_{n,\sigma'}  + {}\nonumber\\
		&c_{n,\sigma}^{\dagger}&\frac{(-\mu_n\delta_{\sigma,\sigma'}+\Gamma \sigma_{\sigma, \sigma'}^z)}{2}c_{n,\sigma'} \Big] + \Delta_n c_{n,\uparrow}^{\dagger}c_{n,\downarrow}^{\dagger} + h.c.\Bigg\},
		\label{eqn:Hamiltonian}
	\end{eqnarray}
where $c^\dagger_{n,\sigma}$($c_{n,\sigma}$) represents  creation (annihilation) operators of fermions with spin $\sigma$ at site $n$. The parameters $h_n$, $\alpha_n$, $\mu_n$, $\Gamma$ and $\Delta$ represent the tunneling, spin-orbit, chemical potential, Zeeman energy and the superconducting terms, respectively at the site $n$. The system under consideration exhibits a characteristic phase transition, between a trivial and a topological phase as the Zeeman term crosses the critical value, $\Gamma_c = \sqrt{\mu^2 + \Delta_0^2}$. The topological phase is accompanied by the appearance of MBS, which are localized at the boundary separating the trivial and the topological region. The movement and manipulation of Majorana bound states (MBS) can be achieved by dynamically shrinking or expanding the topological region through the tuning of control parameters, such as the Zeeman field or the chemical potential.
In this paper we will use the Zeeman term as the control parameter, the qualitative result remain unaffected if instead one used the chemical potential as the control parameter.


 Our work primarily focuses on characterizing dynamical properties, specifically the square overlap between the instantaneous ground state and the time-evolved ground state. This provides a valuable metric to quantify the degree to which the evolving state remains in the ground state configuration. The overlap (also known as the ground-state fidelity) is defined as: 
 \begin{equation} 
 \mathcal{F}(t) = |\langle \psi_{ins}(t) | \psi(t) \rangle |^2. \end{equation}
 The ground-state fidelity is an effective tool for quantifying defect generation in a dynamical setup, offering insights beyond those provided by the square single-particle overlap of in-gap states ~\cite{hegde_quench_2015, truong_optimizing_2023, bauer_dynamics_2018}. We also measure the square single-particle overlap to enable comparison. In this paper, ground-state fidelity will serve as our primary metric.

\section{\label{sec:Piano_s_key}Piano key setup - single key}

In this section, we investigate the dynamics of MBS within the piano key setup. To achieve controlled transport of the MBS by a fixed distance, the region is divided into $ n_k $ segments, or "keys". In a typical piano key setup, each step moves the MBS by a distance corresponding to the length of the key, with the total drive time for the entire movement from the initial to the final position being $ \tau $. The parameters within each key are controlled independently, they are uniformly tuned from an initial value $\Gamma^i$ to a final value $\Gamma^f$ over the time interval given by $\tau/n_k$, enabling precise manipulation of the MBS by sequentially driving segments of the wire into the trivial regime.  When $ \Gamma^f < \Gamma_c $, the corresponding segment transitions into the trivial phase. This process thus facilitates controlled, stepwise movement of the MBS through the entire region.
The baseline parameters for our analysis are as follows: the total system size is $ N = 150 $, with $ h = 25.4 $ meV, $ \alpha = 2 $ meV, and $ \Delta = 0.5 $ meV. The MBS is transported over 48 sites, with the initial and final values of the control parameter set to $ \Gamma^i = 1.4 $ meV and $ \Gamma^f = 0 $ meV, respectively. Unless otherwise specified, these values define the default configuration for our simulations.


\begin{figure}[t]
\centering
\includegraphics[clip=true,width=\columnwidth]{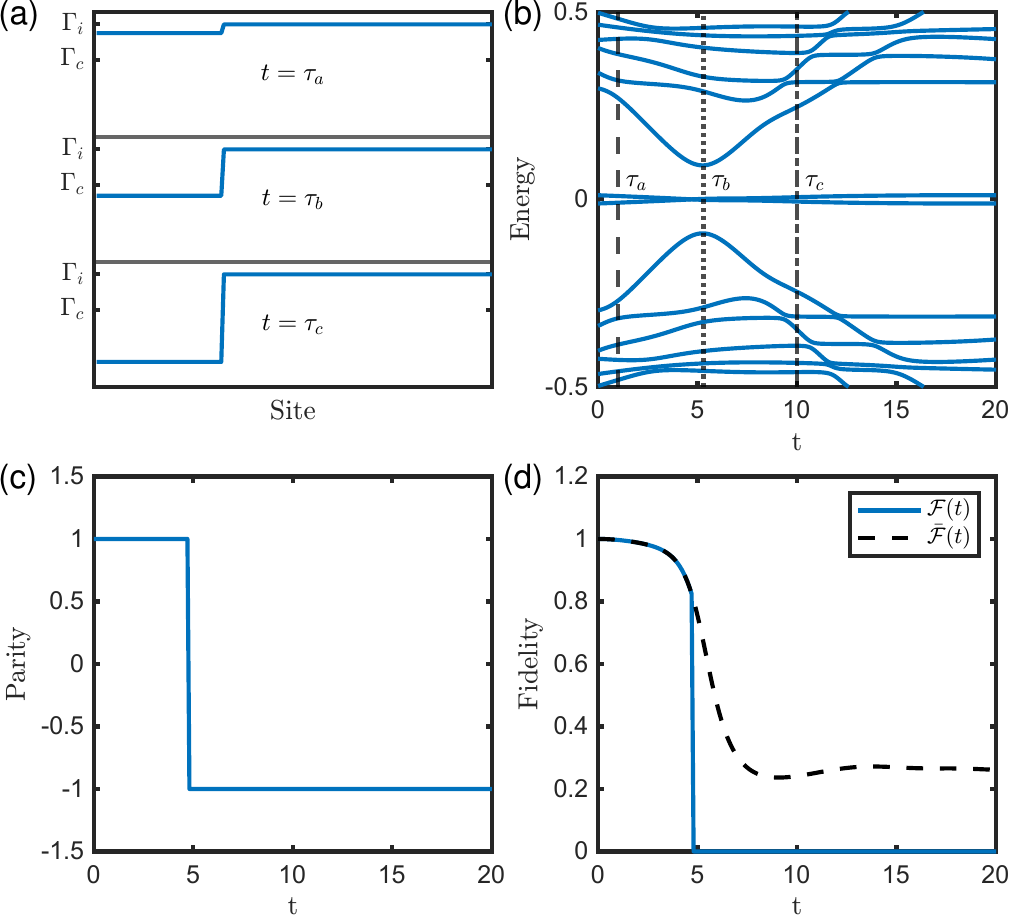}
\caption{\label{fig:piano_spectrum} Piano key setup with single keys for parameter $\tau = 20$ (a) profile depicting key pressing corresponding to different position of key,(b) Energy spectrum with time, (c) parity flipping accompanied by Majorana zero energy crossing, (d) Ground state adiabatic fidelity showing zero due to parity flipping and revised adiabatic fidelity incorporating both parity sectors.}
\end{figure}

We first consider the simplest case involving piano key protocol with a single-key. As the key passes through the critical value (depicted in Fig.~\ref{fig:piano_spectrum}a), the bulk energy spectrum displays features reminiscent of gap closing and reopening, akin to those observed in the topological phase transitions. However, due to the finite length of the wire, the phase transition is local and the  gap is never fully closed. The presence of the topological region ensures that there exist two-states close to zero energy which are localised at the ends of the topological region. Throughout
this protocol, the in-gap modes stay close to zero energy,
with minor modifications in the energy splitting
relative to the gap.
The effect of this single-key setup on the energy spectrum and parity is shown in Fig.~\ref{fig:piano_spectrum}b and
Fig.~\ref{fig:piano_spectrum}c. As the key is pressed the fidelity $\mathcal{F}(t)$  evolves with time (see Fig.~\ref{fig:piano_spectrum}d).  
This dynamical quantity is obtained by evaluating the  following correlator 
\begin{align}
   \bra{\psi_{ins}(t)}  \beta_1 \dots \beta_{2N}& \beta^\dagger_{2N} \dots \beta^\dagger_1 \ket{\psi_{ins}(t)} \CR
    =&|\bra{\psi_{ins}(t)}\ket{\psi(t)}|^2 = \mathcal{F}(t)
\end{align}

where $\beta_i$ are the time-evolved annihilation operator of the initial ground state and $\ket{\psi_{ins}(t)}$ are the instantaneous ground state (see Appendix \ref{app:Paffian}).

The discontinuity in Fig. Fig.~\ref{fig:piano_spectrum}c and Fig.~\ref{fig:piano_spectrum}d are due to the  zero-energy crossing of the Majorana modes, which is accompanied  by the flipping of the  parity  of the instantaneous ground state. This phenomenon, known as parity blocking ~\cite{hegde_quench_2015}, occurs because, starting from an even-parity ground state, it is not possible to reach the odd-parity sector through unitary evolution.  Indeed it is worth noting that the vanishing of fidelity here is not due to the profusion of excitations  but due to the orthogonality of ground states between different parity sectors.  Since our aim is to quantify the excitations generated during the dynamics, we redefine the Ground-state fidelity as the overlap of the time-evolved ground state with respect to  the instantaneous ground state corresponding to each of the parity sectors. This
modified Ground-state fidelity is given by:
\begin{equation}
    \bar{\mathcal{F}}(t) = |\bra{\psi_{ins}(t)}\ket{\psi(t)}|^2 + |\bra{\psi'_{ins}(t)}\ket{\psi(t)}|^2,
\end{equation}
where $\ket{\psi_{ins}(t)}$ and $\ket{\psi'_{ins}(t)}$ are the many-body ground states belonging to different parity sectors with energy  separation of the order of $exp(-L/\xi)$, where $\xi$ is the  the Majorana localization length. We incorporate this definition, to study the effects of drive time duration and noise on the system, irrespective of the parity of the instantaneous ground state. This approach provides a more comprehensive understanding of the dynamics, as illustrated in Fig.~\ref{fig:piano_spectrum}d.

\begin{figure}[t]
    \includegraphics[clip=true,width=\columnwidth]{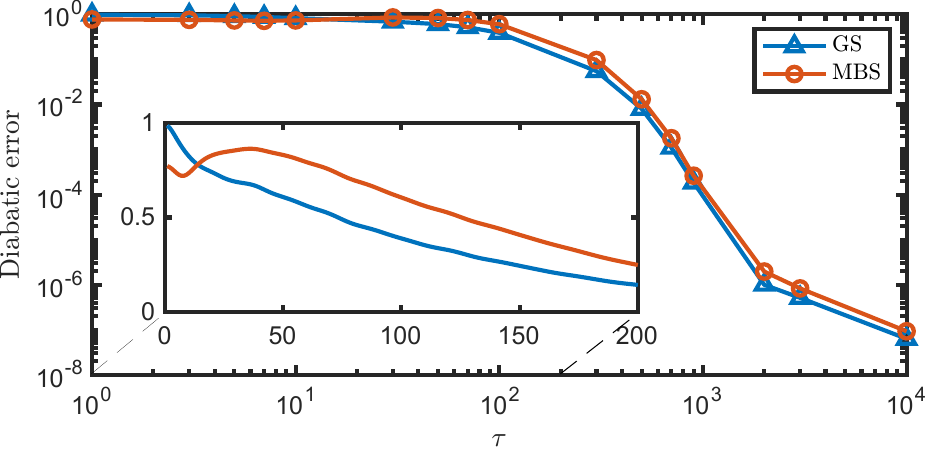}
    \caption{Diabatic error Vs. drive time ($\tau$) for Piano key setup with single keys.}
    \label{fig:piano_fidility}
\end{figure}

The diabatic error, defined as $1-\mathcal{F}(t)$, essentially quantifies the deviation of the system from the instantaneous state. This measure will be used to assess how well the system remains in its ground state during the evolution. With these definitions in place, we examine the effect of increasing drive time, $\tau$, on Diabatic error (see Fig.~\ref{fig:piano_fidility}). Diabatic error exhibits monotonic behavior as a function of the drive time. Similar to the noiseless case for spinless systems ~\cite{truong_optimizing_2023,bauer_dynamics_2018}, we demonstrate that in the spinful case, the diabatic error exhibits an exponential decay for small drive times, transitioning to a power-law decay for large drive times. As the drive time is increased, the final state remains nearly identical to the instantaneous ground state at the end of the drive resulting in suppressed diabatic errors. It turns out that besides the ground-state diabatic error, the single-state diabatic error of the MBS defined as $1-\mathcal{M}(t)$ with $\mathcal{M}(t) = |\bra{0} b_{2N}(t) \beta_{2N}^\dagger(t) \ket{0}|^2$, captures the overall qualitative behavior  and can serve as a valid measure in certain regimes. However, under fast driving, the measure deviates significantly from the ground-state diabatic error, suggesting it becomes less reliable in capturing the system's overall deviation from the intended adiabatic evolution. In many applications, such as quantum annealing or adiabatic quantum computation, the primary objective is to maintain the system as close as possible to its ground state throughout the process. Consequently, the ground-state diabatic error emerges as a more meaningful metric for assessing the system's adherence to adiabatic behavior.

\subsection{\label{subsec:s_key_noise} Effect of noise}

To accurately model realistic conditions in dynamical scenarios involving the manipulation of quantum states,
it is crucial to account for and understand the effect of noise on the  defect/error production.
In this subsection, we analyze how noise, modeled as a perturbation to the Zeeman term, affects the adiabatic fidelity of the ground state. Additionally, we investigate how various noise characteristics influence adiabaticity.

We define the dichotomous noise $ \eta(t) $ as a two-state stochastic process oscillating between values $ a $ and $ b $ with switching rates $ \mu_a $ and $ \mu_b $, respectively. The stationary probabilities are $ P_s(a) = \mu_b/(\mu_a + \mu_b) $ and $ P_s(b) = \mu_a/(\mu_a + \mu_b)$, while the stationary mean and correlation conditions are given by:
\begin{equation}
\label{eq:eta}
    \langle \eta(t) \rangle_s = \frac{a \mu_b + b \mu_a}{\mu_a + \mu_b}, \quad \langle \eta(t) \eta(t') \rangle = \sigma^2 e^{-|t - t'| / \tau_c},
\end{equation}
where $ \tau_c = 1 / (\mu_a + \mu_b) $ and $ \sigma^2 = \frac{(a - b)^2 \mu_a \mu_b}{(\mu_a + \mu_b)^2} $. We ensure $ \langle \eta(t) \rangle = 0 $ by setting $ a \mu_b + b \mu_a = 0 $.

\begin{figure}[t]
\centering
\includegraphics[clip=true,width=\columnwidth]{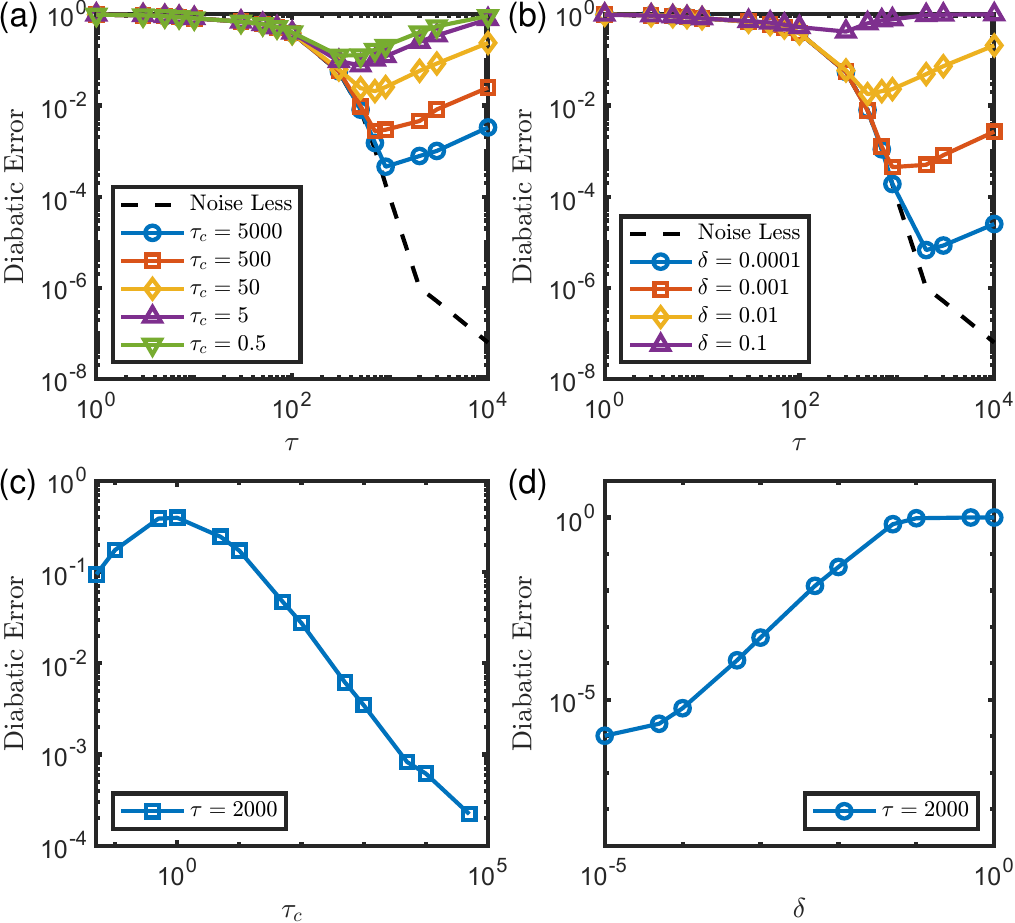}
\caption{Effect of noise parameters on diabatic error. (a) effect of noise correlation time on $\tau$ vs. diabatic error, for various correlation times. (b) Influence of noise strength on $\tau$ vs. diabatic error, for different noise strengths. (c) Variation of diabatic error with noise correlation time for a fixed $\tau$. (d) Variation of diabatic error with noise strength for a fixed $\tau$.}
\label{fig:noisy_piano_single}
\end{figure}

The conditional probabilities $ P_{aa} $ and $ P_{bb} $ are obtained based on the switching rates $ \mu_a $ and $ \mu_b $.
further details regarding the noise protocol along with the algorithm used to generate the dichotomous noise sequence $ \eta(t) $ is provided in Appendix \ref{app:noise}. We use symmetric dichotomous noise with parameters $ a = -\delta $, $ b = \delta $, $ \mu_a = \mu_b = k_0 $, yielding $ \tau_c = 1 / (2k_0) $. We explore how different aspects of the noise (present in Zeeman term) affect diabatic error such as the noise strength, $\delta$, and the noise correlation time, $\tau_c$ as depicted in Fig.~\ref{fig:noisy_piano_single}. 
The effect of noise on diabatic error depends strongly on the drive time. For extended drive times, diabatic errors increase significantly as the system parameter (the Zeeman term, within the region where the piano key is pressed) remains near the critical point for longer durations, amplifying the effect of noise and consequently enhancing defect production. This enhanced defect generation results in a pronounced deviation from the noiseless case. In contrast, for shorter drive times, the system spends less time near critical points, reducing the influence of noise and leading to minimal deviation from the noiseless case. This behavior highlights the existence of an optimal drive time that minimizes the impact of noise for a given correlation time.

For large noise correlation times, the deviation from the noiseless case is minimal. The system can closely follow the noiseless behavior for longer drive times before the effects of noise become significant. However as the noise correlation time decreases, the effect of noise becomes evident at shorter drive times, as shown in Fig.~\ref{fig:noisy_piano_single}a.  To gain deeper insight into this behavior, we examine different noise correlation times  for a fixed drive time. The diabatic error initially increases as the noise correlation time decreases, reaching a maximum before starting to decline, as shown in Fig.~\ref{fig:noisy_piano_single}c. This behavior aligns with the following expectations: for large noise correlation times, infrequent noise fluctuations have a reduced effect, while for very small correlation times, rapid fluctuations average out the noise contribution, rendering their effects less significant. The influence of noise strength on diabatic error is illustrated in Fig.~\ref{fig:noisy_piano_single}b. As expected, decreasing the noise strength allows the system to adhere to the noiseless case for longer drive times before the diabatic error becomes significant. This behavior is further confirmed when analyzing diabatic error at a fixed drive time (see Fig.~\ref{fig:noisy_piano_single}d). For very high noise strength, the diabatic error approaches its maximum value, which in this case is nearly 1.

\subsection{Effective model and Scaling\label{subsec:s_key_effective}}
In this section, we derive an effective two-level model for the SM-SC heterostructure with a single piano key to predict error scaling for both the  noiseless and noisy drive scenarios. Our approach builds on the works of Bauer et al.~\cite{bauer_dynamics_2018} and Truong et al.~\cite{truong_optimizing_2023}, in that we model the problem in terms of an effective Landau-Zener Hamiltonian taking into consideration  the following points.

First, in relatively slow driving scenarios, the major contribution to the diabatic error arises from excitations from the exponentially separated low lying many-body state to the nearest excited state with the same parity. Second, the probability of such transitions is highest when the key passes through the critical parameter. 
When the key reaches criticality, the Majorana modes (which was at the edge touching the key) and the lowest bulk modes become localized within the key region. Therefore, under this scenario, one can treat the key as a separate chain and use this to estimate the bulk gap~\cite{bauer_dynamics_2018,truong_optimizing_2023}. However, the bulk mode leaks partially into  the topological region and therefore the estimation of the gap requires certain modifications as is discussed below.

Consider the Hamiltonian given by Eq.~\eqref{eqn:Hamiltonian} in the Nambu basis, $\Psi_n = (c_{n\uparrow}, c_{n\downarrow}, c^\dagger_{n\downarrow}, -c^\dagger_{n\uparrow})^{T}$:
\begin{equation}
\label{BDG}
H_{\text{BDG}} = \frac{1}{2} \sum_n \left[ \Psi^{\dagger}_n \hat{A}_n \Psi_n + (\Psi^{\dagger}_n \hat{B}_n \Psi_{n+1} + \text{h.c.}) \right],
\end{equation}
where the matrices $\hat{A}_n$ and $\hat{B}_n$ are defined as:
\begin{align}
\hat{A}_n &= - \mu_n \sigma_0 \tau_z + \Gamma \sigma_z \tau_0 + \Delta_n \sigma_0 \tau_x,\\
\hat{B}_n &= -h \sigma_0 \tau_z -i \alpha \sigma_y \tau_z.
\end{align}
Here, $\sigma_i$ and $\tau_i$ are the pauli matrices acting on the spin and particle-hole spaces, respectively. To find the bulk bands it is convenient to consider the following momentum space Hamiltonian,
\begin{eqnarray}
H(k) = \frac{1}{2}\left[ -\mu \tau_z \otimes \sigma_0 + \Gamma \tau_0 \otimes \sigma_z + \Delta \tau_x \otimes \sigma_0 \right.\nonumber\\
\left. -2h \cos(k) \tau_z \otimes \sigma_0 + 2\alpha \sin(k) \tau_z \otimes \sigma_y \right].
\end{eqnarray}
The bulk gap closes at momentum $k=0$.
Expanding the Hamiltonian around $k =0$ and redefining the chemical potential relative to the bottom of the band, i.e., $\mu \rightarrow \mu -2h$ we obtain:

\begin{eqnarray}
H(k) = \frac{1}{2}\left[ -\mu \tau_z \otimes \sigma_0 + \Gamma \tau_0 \otimes \sigma_z + \Delta \tau_x \otimes \sigma_0 \right.\nonumber\\
\left. + k^2 h \tau_z \otimes \sigma_0 + 2\alpha k \tau_z \otimes \sigma_y \right].
\end{eqnarray}

We next focus on the eigenvalues close to zero energy. After some simplification, the energy spectrum can be approximated as:
\begin{equation}
\epsilon \approx \pm \sqrt{(\Gamma-\Gamma_1)^2 + 4 \alpha^2 \bar{k}^2 \left(1-\frac{(\mu - \bar{k}^2 h)^2}{\Gamma_1^2}\right)},
\end{equation}
where $\bar{k} = \pi/L_{key}$ which is momentum resolution and $\Gamma_1^2 = \Delta^2 + (\mu - \bar{k}^2 h)^2$ is the modified critical zeeman term for the finite key length which goes to $\Gamma_c$ in the limit $L_{key} \rightarrow \infty$.

This allows us to write an effective two-level Landau-Zener problem as:
\begin{equation}
    H_{\text{eff}}(t) = \frac{1}{2} \begin{pmatrix}
        \Gamma(t) - \Gamma_1 & \Delta_f \\
        \Delta_f & -(\Gamma(t) - \Gamma_1)
    \end{pmatrix},
\end{equation}
where
\begin{equation}
\Delta_f = \sqrt{4 \alpha^2 \bar{k}^2 \left(1 - \frac{(\mu - \bar{k}^2 h)^2}{\Gamma_1^2}\right)},
\end{equation}
and
\begin{equation}
\Gamma(t)= \left(1-\frac{t}{\tau}\right) \Gamma_i + \frac{t}{\tau} \Gamma_f.
\end{equation}

\begin{figure}[tp]
\centering
\includegraphics[width=\columnwidth]{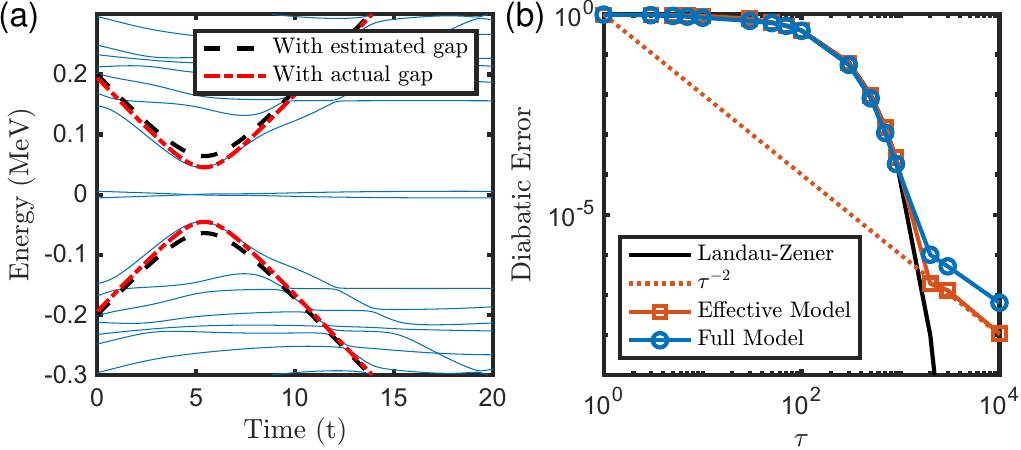}
\caption{Comparison between the effective and full models for a system with given parameters. (a) Energy spectrum comparison: the effective model is evaluated using the estimated gap ($\Delta_f$) and the actual gap ($\Delta_a$), shown together with the full model spectrum. (b) Diabatic error comparison: both models agree well in exhibiting exponential scaling in the short drive-time regime, transitioning to a power-law decay of $\tau^{-2}$ in the long drive-time limit, though the effective model underestimates the power-law contribution.}
\label{fig:full_comp}
\end{figure}

As discussed earlier, in our system, the bulk modes leak into the adjacent regions, causing the gap to be overestimated, as shown in Fig.~\ref{fig:full_comp}a. Since our main aim is to develop an effective low-energy model for analytical calculations, we address this issue by using the actual gap, $\Delta_a$, in place of $\Delta_f$, a similar approach was also considered in ~\cite{truong_optimizing_2023}. With this modification, we have an effective low-energy model that describes the system well around the band-closing point, as illustrated in Fig.~\ref{fig:full_comp}a.

The diabatic error for the case of a two-level problem driven from time $-\infty$ to $\infty$ is given by the Landau-Zener formula as:
\begin{equation}
    D_{LZ} = \exp\left(-\frac{\pi \Delta_a^2}{2\dot{\Gamma}}\right).
\end{equation}

In the presence of a finite drive time $\tau$, the diabatic error exhibits a power-law contribution that becomes more pronounced for slow driving, as shown in Fig.~\ref{fig:full_comp}b. The effective model captures the behavior of the diabatic error for the full model quite well, particularly the exponential suppression of diabatic transitions. However, for longer drive times, the effective model slightly underestimates the power-law contribution. Nonetheless, it still captures the overall scaling behavior, which is proportional to $\tau^{-2}$. Thus, despite this slight underestimation, the effective model is able to capture the power-law scaling of the diabatic error observed in the full model.

Next, we consider drive in the presence of noisy fluctuations in the Zeeman term. Our aim in this section is to incorporate the noise term in the effective model and determine the modifications to the diabatic error arising due to this additional term and compare the results with those obtained from the full model.

\begin{figure}[t]
\centering
\includegraphics[width=\columnwidth]{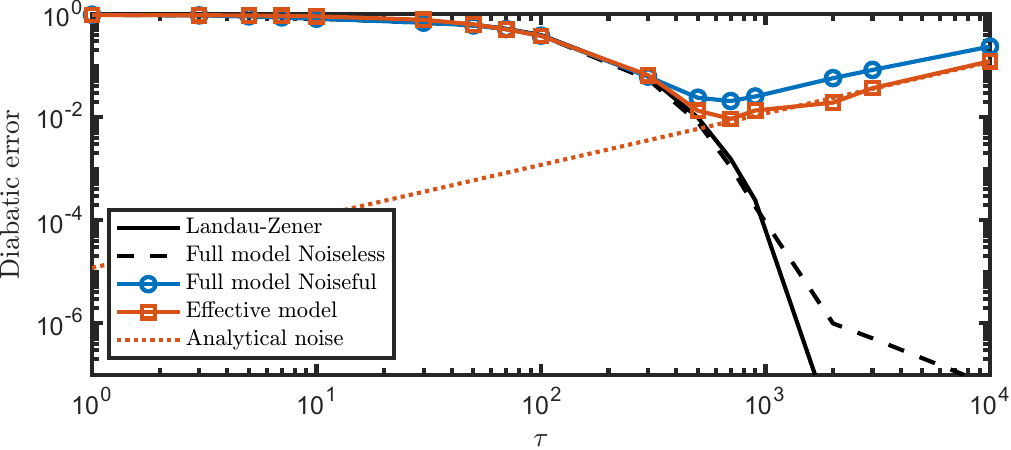}
\caption{Comparison of the diabatic error in the presence of noise for the effective and full models.}
\label{fig:noise_comp}
\end{figure}

The effective model in the presence of noise is given by:
\begin{equation}
    H_{\text{eff}}(t) = \frac{1}{2} \begin{pmatrix}
        \Gamma(t) + \eta(t) - \Gamma_1 & \Delta_a \\
        \Delta_a & -(\Gamma(t) + \eta(t) - \Gamma_1)
    \end{pmatrix},
\end{equation}
where $\eta(t)$ is the telegraph noise term as defined in Eq.~\ref{eq:eta}. The two-level Landau-Zener problem in the presence of noise has garnered significant attention, with perturbative solutions developed by Malla et al. and Krzywda et al. In our work, we use the results obtained by Krzywda et al. to predict the behavior of diabatic error under noisy conditions in the long-time limit and compare these predictions to numerical results from both the effective and full models. The predicted excitation probability for a specific two-level system due to the telegraph noise in the slow driving regime is given by:
\begin{equation}
    D_{noise} = \tau\frac{\sigma^2 \pi \tau_c \Delta_a }{2(\Gamma_f - \Gamma_i)}\left(1 - \frac{1}{\sqrt{1 + \frac{1}{\Delta_a^2 \tau_c^2}}}\right).
\end{equation}

The comparison between the full model and the effective model is illustrated in Fig.~\ref{fig:noise_comp}. The effect of noise is particularly evident in the slow driving case, where the behavior of the diabatic error changes significantly. In the presence of noise, after an optimal drive time, the diabatic error begins to increase, following the predicted linear scaling in $\tau$. Both the effective model and the full model exhibit the same scaling in diabatic error. However, the effective model slightly underestimates the diabatic error.

This underestimation arises primarily because the effective model only considers the interaction between two states. In contrast, the full model encompasses a significantly larger Hilbert space, allowing for more possible transitions and interactions. This broader space leads to more pathways for diabatic transitions, resulting in a comparably higher diabatic error in the full model. For this particular reason, the diabatic error saturates at $0.5$ for the effective model, while it is close to $1$ for the full model.

\begin{figure}[t]
    
\centering
    \includegraphics[clip=true, width=\columnwidth]{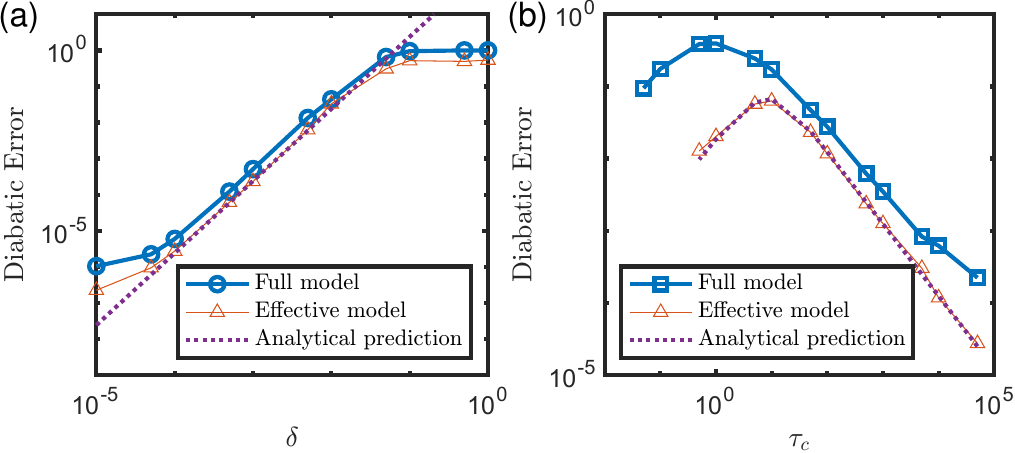}
    \caption{Effect of noise parameters on diabatic error for both the effective and full models, along with analytical predictions for the effective model. (a) Diabatic error as a function of noise correlation time for a fixed $\tau$. (b) Diabatic error as a function of noise strength for a fixed $\tau$. Despite slightly underestimating the diabatic error, the effective model captures the qualitative behavior well.}
    \label{fig:noise_main}
\end{figure}

Using the effective model, we can further explore how various noise characteristics influence the diabatic error and compare these results with those from the full model. The behavior of the diabatic error as a function of noise strength for a fixed drive time is shown in Fig.~\ref{fig:noise_main}a. As the noise strength increases, the diabatic error reaches a maximum value $0.5$ in the effective model and nearly $1$ in the full model. Focusing on the effect of noise correlation time for a specific drive time, we observe that as the noise correlation time decreases, the diabatic error initially rises, peaking at a certain correlation time before decreasing again, as illustrated in Fig.~\ref{fig:noise_main}b. Despite the effective model underestimating the overall error, it successfully captures the qualitative trends in diabatic error with respect to both noise strength and noise correlation time. Moreover, the scaling behavior observed in the effective model aligns well with analytical predictions, as demonstrated in Fig.~\ref{fig:noise_main}.

\begin{figure*}[t]
\centering
\includegraphics[clip=true,width=\textwidth]{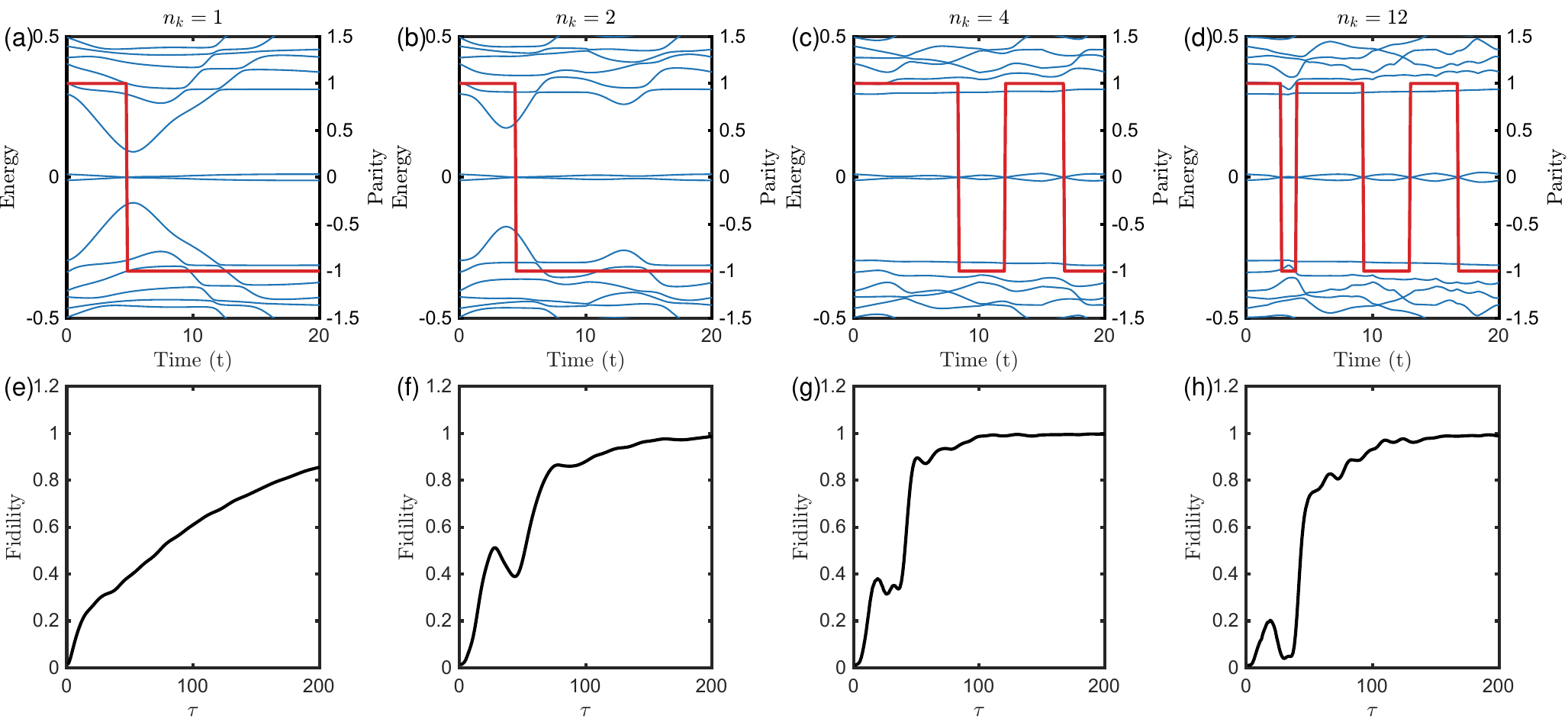}
		\caption{Time evolution of the energy spectrum and parity for different numbers of keys in a piano key setup, alongside ground-state and MBS fidelity for a range of drive times ($\tau$).
(a)-(d): Time evolution of the energy spectrum (red) and parity (blue) for four different numbers of keys: 1, 2, 4, and 12, respectively. The energy spectrum is plotted on the left y-axis, while the parity is plotted on the right y-axis. Parity flipping and Majorana oscillations increase as the number of keys increases.
(e)-(h): Ground-state fidelity $\bar{\mathcal{F}}(\tau)$ vs. drive time ($\tau$) for different key values in the same systems, showing the dynamical behavior of these quantities. As the number of zero energy crossings increases with the number of keys, the Landau-Zener-Stückelberg (LZS) interference also increases, causing more oscillations in fidelity.}

		\label{fig:piano_n_key}
	\end{figure*}
    
\section{Piano key setup - Multiple key\label{sec:piano_m_key}}
In this section we explore the effect of multiple keys on the energy spectrum, parity flips, and fidelity (for similar drive protocol as discussed above). As the number of keys increases, we observe significant deviations in the behavior of the bulk bands. For a small number of keys, the bulk band shows marked dips in the energy spectrum, but as the number of keys increases, the effect on the bulk bands becomes less pronounced. Even with 4 keys, the bulk gap remains mostly unchanged.
This is due to the fact that as the number of keys is increased only a small segment of the wire passes through critical point while the majority of the segment remain in the non-critical regime.
This is amply demonstrated in Figs.~\ref{fig:piano_n_key}a-d, where the energy spectrum is plotted for the system with 1, 2, 4, and 12 keys, respectively. It is also clear from the figures that the number of parity flips and the MBS oscillations increases with the number of keys. This indicates that both the complexity of the protocol and the number of keys significantly influence the dynamics of the system.

To further analyze these effects, in Figs~\ref{fig:piano_n_key}e-h we present the ground-state fidelity $\bar{\mathcal{F}}(\tau)$ as function of the drive time, $\tau$, for the same key configurations. In addition to the expected increase in fidelity with longer drive times, we also observe oscillations in the fidelity values. These oscillations can be attributed to the Landau-Zener-Stückelberg (LZS) interference, which occurs when the energy levels of the system undergo crossings with very small gaps~\cite{xu_dynamics_2023}. During such crossings, there is a probability for the system to make non-adiabatic transitions between states, leading to interference effects that manifest as oscillations in the fidelity.
As the number of keys increases, the number of zero energy crossings and parity flips rises, leading to stronger Landau-Zener-Stückelberg (LZS) interference. This interference manifests as more pronounced oscillations in the fidelities, thereby highlighting the intricate relationship between zero energy crossings and parity flipping on archiving a certain state.

\subsection{Optimal number of key\label{subsec:optimal_key}}
In this section, we address the question regarding the optimal number of keys by considering both the noiseless and noisy scenarios. Interestingly the answer differs for the two scenarios and therefore has implications for the experimental realization of a moving Majorana setup.

\begin{figure}[t]
\centering
\includegraphics[width=\columnwidth]{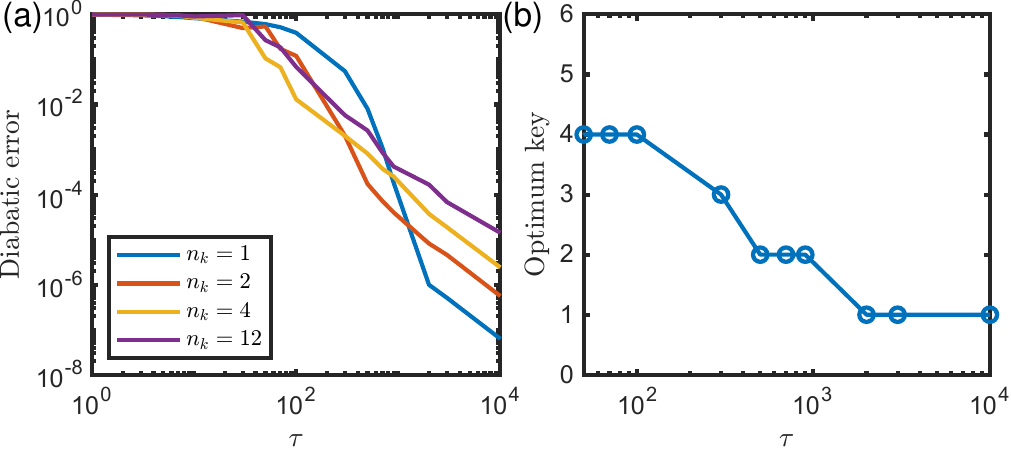}
\caption{(a) Effect of the number of keys on the diabatic error without any noise. (b) Optimal number of keys for different drive times. In the intermediate drive time regime, using multiple keys results in a lower diabatic error.}
\label{fig:optimal_1}
\end{figure}

Figure~\ref{fig:optimal_1}a illustrates the diabatic error as a function of drive time for different numbers of keys. In Figure~\ref{fig:optimal_1}b, we plot the optimal number of keys versus drive time. 
In the noise-free case, we observe that a single key performs better in the slow driving regime. However, in the intermediate regime, where diabatic error is  around $10^{-2}$ which is still acceptable range, using more keys results in marginal improvement in performance. In our setup, we find that four keys, and in some cases two keys, are optimal.

\begin{figure}[t]
\centering
\includegraphics[width=\columnwidth]{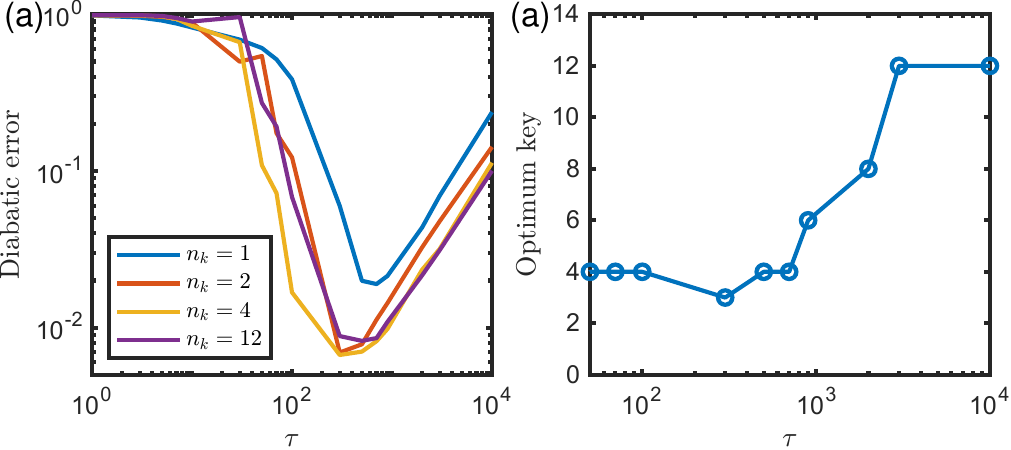}
\caption{(a) Effect of the number of keys on the diabatic error in a noisy scenario. (b) Optimal number of keys for different drive times. In the noisy case, a single key is no longer optimal, and increasing the number of keys results in better performance.}
\label{fig:optimal_2}
\end{figure}

The picture changes when noise is introduced, as shown in Figs.~\ref{fig:optimal_2}. In the presence of noise, a single key becomes less effective in the slow-driving regime, whereas increasing the number of keys helps reduce diabatic errors. Notably, in the very slow driving case, using twelve keys results in the lowest diabatic error, as shown in Fig.~\ref{fig:optimal_2}b. However, Fig.~\ref{fig:optimal_2}a shows that the the reduction in error as compared to that of the  4 key arrangement  is insignificant. This is an important consideration, as increasing the number of keys complicates the experimental setup. Therefore, if the improvement is minimal, it is preferable to maintain a simpler setup with fewer number of keys. This result is for a particular noise parameters. Similar analysis has been performed for different noise correlation time and strength and we find the qualitative pictures remains the same. To summarize, as opposed to the noiseless scenario for noisy scenario multiple
key performs better then the single key.

\section{Disorder and Inhomogeneity\label{sec:disorder}}
In this section, we investigate the influence of disorder and inhomogeneity on the transport dynamics of Majorana bound states (MBS). Such perturbations are intrinsic to realistic physical systems and can significantly impact the robustness of MBS transport. The focus of our analysis is again to evaluate the extent to which the time-evolved quantum state adheres to the instantaneous ground state of the system during the transport process. 

\begin{figure}[t]
\centering
\includegraphics[width=\columnwidth]{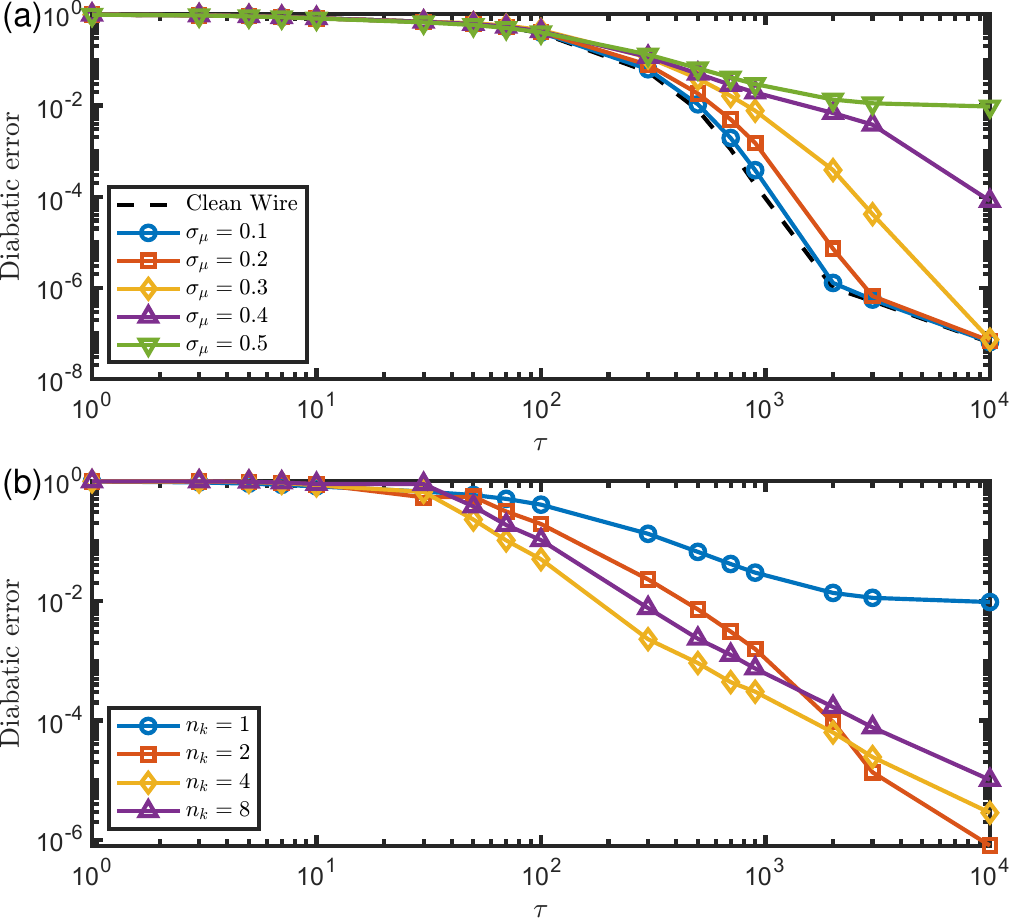}
\caption{(a) Diabatic error as a function of drive time for various disorder strengths $\sigma_\mu$. (b) Average diabatic error for different numbers of keys in a disordered nanowire with  $\sigma_\mu = 0.5$.}
\label{fig:disorder}
\end{figure}

\begin{figure}[t]
\centering
\includegraphics[width=\columnwidth]{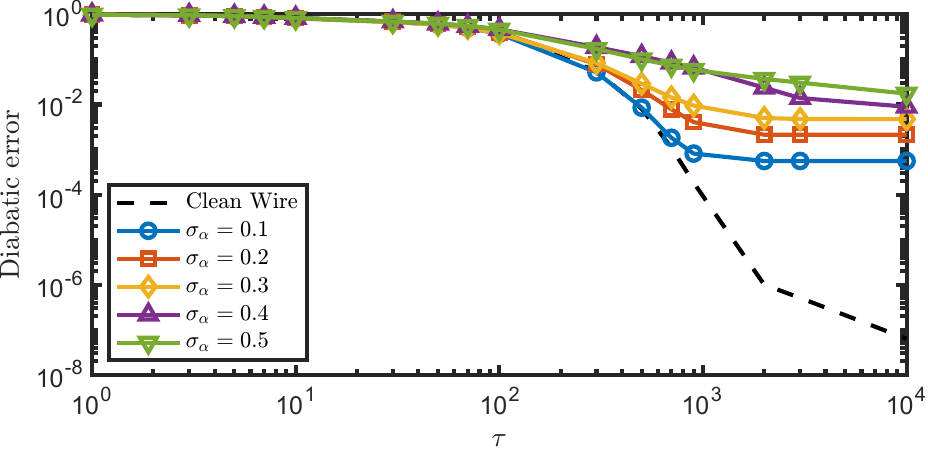}
\caption{effect of spin-orbit coupling disorder on the diabatic error as a function of drive time for various disorder strengths $\sigma_\alpha$.}
\label{fig:disorder_alpha}
\end{figure}

We first consider random Gaussian disorder on top of the fixed value of chemical potential. The disorder is characterized by zero mean and standard deviation, $\sigma_\mu$. The diabatic error associated with the transport of a Majorana mode using a single key in a disordered system (averaged over multiple realizations) is presented in Fig..~\ref{fig:disorder}a. 
With the increase of disorder strength we find that  the averaged diabatic error exhibits an increase for all drive times, with the reduction in the diabatic error happening  more slowly  (with respect to the drive time)  compared to that for the clean case.   We further investigated the effect of the number of keys within a specific range of disorder. Our findings indicate that the average error improves when using multiple keys compared to a single key (see Fig.~\ref{fig:disorder}b). 
The second scenario we considered involves in the spin-orbit coupling term. Specifically, we added a disordered spin-orbit coupling component in the $x$-direction (normal to the uniform spin-orbit coupling term), with the disorder characterized by a zero mean and a standard deviation $ \sigma_\alpha $. In the presence of this disorder we obtain similar behavior as for the disorder in the chemical potential, i.e., the average diabatic error increases and deviates from the behavior observed in the clean case as shown in Fig.~\ref{fig:disorder_alpha}.

\begin{figure}[t]
\centering
\includegraphics[width=\columnwidth]{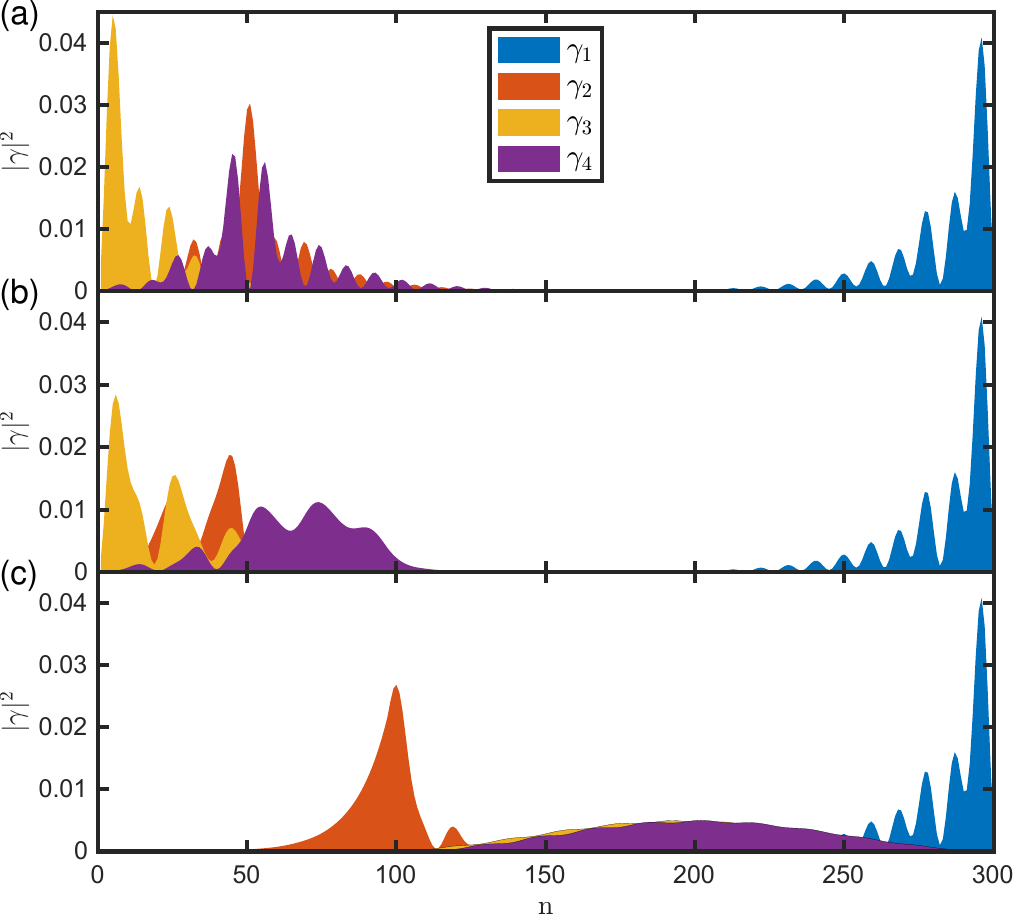}
\caption{Majorana Bound States during the piano key protocol with a kink at site 50. (a) Initial configuration with the key unpressed. (b) Configuration as the key is pressed up to the critical parameter. (c) Final configuration at the end of the protocol.}
\label{fig:kink_MBS}
\end{figure}

\begin{figure}[t]
\centering
\includegraphics[width=\columnwidth]{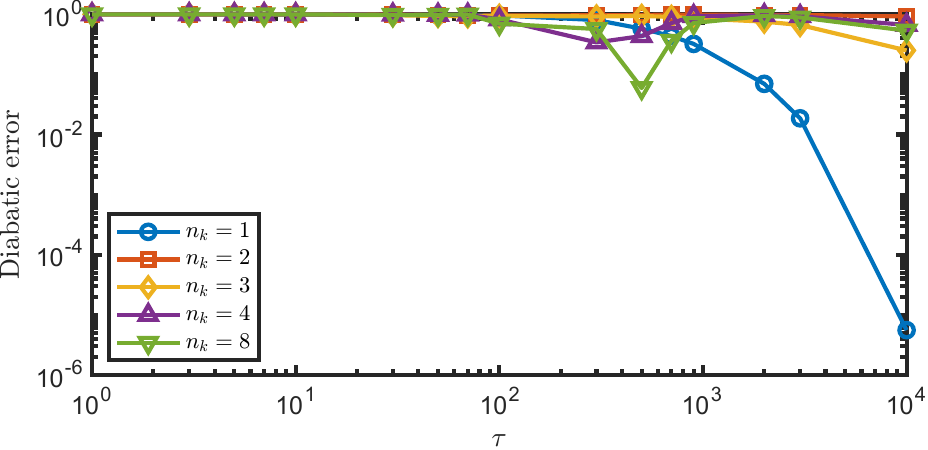}
\caption{Effect of number of keys on diabetic error scaling with drive time}
\label{fig:kink_50}
\end{figure}

Lastly, we examined a type of inhomogeneity, which can alter the system more drastically than disorder considered above. Specifically, This involves a kink in the spin-orbit coupling term (reversal in the coupling direction)
which effectively divides the nanowire into two regions each of which having  homogeneous parameters. We considered this case in a 300 site chain with kink at 50th position, in the topological parameter space, this configuration results in four MBS (see Fig.~\ref{fig:kink_MBS}a). Before delving  into behavior of diabetic error, let us first analyze the behavior of instantaneous states by defining four majorana states from four low lying modes (where $\gamma_1$ and $\gamma_2$ belong to the right part of the wire while $\gamma_3$ and $\gamma_4$, belong to the right part).The dynamics involving the transport of the Majorana mode $\gamma_3$ across the kink now becomes extremely nontrivial. The mode $\gamma_3$ begins to move and overlaps with its pair $\gamma_4$. As the key approaches (here the single key length is such that it  encapsulates the left 3 Majorana modes) critical parameter all three states in the key i.e. $\gamma_3$, $\gamma_4$ and $\gamma_2$ become delocalized within the key (see Fig.~\ref{fig:kink_MBS}b). Finally when the key enters the trivial parameter regime $\gamma_3$ and $\gamma_4$ overlap completely and move to bulk, while $\gamma_2$ relocates to the edge of the  key (see Fig.~\ref{fig:kink_MBS}c). This dynamics introduces significant complexity when we try to explore the scaling of diabatic error with drive time using different number of keys. As illustrated in Fig.~\ref{fig:kink_50}, increasing the number of keys in this setup generally increases the diabatic error instead of reducing it. This counterintuitive result arises because multiple keys introduce additional zero-energy crossings, which hinder adiabatic evolution.

\section{Conclusion\label{sec:conclusion}}

In this study, we investigated the transport of Majorana Bound States (MBS) in a semiconductor-superconductor heterostructure using a piano key setup, focusing on diabatic error as a function of drive time. To evaluate the system's ability to remain in its ground-state configuration during evolution, we used the dynamic square overlap between the instantaneous and time-evolved ground states. Our analysis showed that the piano key setup induces zero-energy crossings of Majorana modes, accompanied by parity flips, which complicate the study of system dynamics. To address this, we redefined ground-state fidelity as the overlap between the time-evolved ground state and the instantaneous ground state with the correct parity. This adjustment provided reliable metrics for examining the system's behavior under varying drive times and noise conditions. We observed that diabatic error exhibits an initial exponential decrease with increasing drive time, followed by a power-law decay. This behavior underscores the importance of optimizing drive time to minimize errors, particularly in applications such as quantum annealing and adiabatic quantum computation, where maintaining the system in its ground state is crucial.

To simulate realistic scenarios, we incorporated symmetric dichotomous temporal noise into the drive and found that noise significantly affects diabatic error depending on drive time. Longer drive times near critical points amplified defect production, while shorter drive times reduced noise influence, resulting in minimal deviations from the noiseless case. This highlighted the existence of an optimal drive time that minimizes the effect of noise for a given correlation time. We developed an effective model to predict error scaling in both noisy and noiseless scenarios, providing valuable insights into the interplay between drive time, noise, and error production. Extending our analysis, we studied the effect of increasing the number of piano keys, finding that while additional keys do not always improve performance, an optimal number exists for a range of drive times. Specifically, in noiseless, slow-driving regimes, a single key achieves high fidelity, whereas under noisy conditions, additional keys are beneficial, with four keys emerging as optimal in our setup. Finally, we examined the effects of disorder and inhomogeneity on error production during the transport of Majorana modes. Both factors increase the average error and introduce additional complexities, further complicating the dynamics of Majorana transport. These findings provide a comprehensive understanding of the factors influencing error production in MBS transport and offer guidance for optimizing adiabatic protocols in realistic scenarios.

\begin{acknowledgments}
		We thank Professor Daniel Loss for many illuminating discussions and for the hospitality extended to S.G. during the visit to Basel. D.S. would like to thank CSIR for funding by their file
no. 09/1020(0216)/2021-EMR-I.
	\end{acknowledgments}
\appendix
\section{Time evolution and Paffian Formalism}
\label{app:Paffian}
we start by considering the Hamiltonian in BdG basis: 
 \begin{equation}
     H(t) = \tilde{C}^\dagger \mathcal{H}(t) \tilde{C}
 \end{equation}
where $\tilde{C} = \left(c_{1\uparrow}, c_{1\downarrow}, c_{1\downarrow}^\dagger, -c_{1\uparrow}^\dagger, \dots\right)^T$ is a vector of dimension $4N \times 1$, and $\mathcal{H}(t)$ is a $4N \times 4N$ matrix that explicitly incorporates the time dependence of the Hamiltonian.

By diagonalizing the Hamiltonian, we obtain:
 \begin{equation}
    H(t) = \tilde{b}^\dagger(t) \lambda(t) \tilde{b}(t)
 \end{equation}
 where $\lambda(t)$ the diagonal matrix containing eigenvalues and $\tilde{b} = \left(b_1,\dots,b_{2N},b_{2N}^\dagger,\dots,b_1^\dagger\right)^T$ connected to $\tilde{C}$ by $\tilde{b}(t) = B(t)\tilde{C}$ with $B$ containing eigenvectors in rows. The initial ground state satisfies $b_i(0) \ket{\psi(0)} = 0$, while the instantaneous ground state satisfies $b_i(t) \ket{\psi_{ins}(t)} = 0$. We also define the time-evolved ground state, which has its own annihilation operator satisfying $\beta_i(t) \ket{\psi(t)} = 0$, where $\beta_i(t)$ is the time-evolved form of $b_i(0)$.

We consider a stepwise unitary evolution of the state $ \tilde{\beta}(t) = \bar{B}(t)\tilde{C} $ at time $ t $. After a small time increment $ \Delta t $, the state evolves as:
\begin{equation*}
\tilde{\beta}(t+\Delta t) = U \tilde{\beta}(t) U^\dagger = \bar{B}(t) U B^{-1}(t) b(t) U^\dagger.
\end{equation*}
Since the operator $ b(t) $ gains a phase factor under unitary evolution, we have:
\begin{equation*}
U b(t) U^\dagger = e^{-i \lambda(t) \Delta t} \tilde{b}(t),
\end{equation*}
which simplifies the expression to:
\begin{equation*}
\tilde{\beta}(t+\Delta t) = \bar{B}(t) B^{-1}(t) e^{-i \lambda(t) \Delta t} \tilde{b}(t).
\end{equation*}
Thus, the time evolution of $ \bar{B}(t) $ over the interval $ \Delta t $ is compactly written as:
\begin{equation*}
\bar{B}(t+\Delta t) = \bar{B}(t) B^{-1}(t) e^{-i \lambda(t) \Delta t} B(t).
\end{equation*}
Given this, the full time evolution can be calculated by starting from the initial condition which is  $\bar{B}(0) = B(0)$. Next, we consider the following quantity:
\begin{align}
    &\bra{\psi_{ins}(t)} \beta_1 \dots \beta_{2N} \beta^\dagger_{2N} \dots \beta^\dagger_1 \ket{\psi_{ins}(t)} \CR
    =&\bra{\psi_{ins}(t)} \beta_1 \dots \beta_{2N} \beta^\dagger_{2N} \dots \beta^\dagger_1 \sum_{a=1}^{2^{2N}}\ket{\phi_a}\bra{\phi_a}\ket{\psi_{ins}(t)} \CR
    =&|\bra{\psi_{ins}(t)}\ket{\psi(t)}|^2 = \mathcal{F}(t)
\end{align}
where we have inserted the identity operator in terms of the full set of orthonormal states. Note that $|\phi_1\rangle= |\psi(t)\rangle$, while $|\phi_a\rangle$ are excited states annihilated by $\beta^\dagger_i$. For further simplification one makes use of the Paffian formalism to calculate the expectation value of any string of operators by utilizing the property of wicks theorem. 
In our case we want to calculate the square of the overlap between instantaneous ground state and time evolved ground state which is equivalent to  expectational value $\bra{\psi_{ins}(t)} \beta_1 \dots \beta_{2N} \beta^\dagger_{2N} \dots \beta^\dagger_1 \ket{\psi_{ins}(t)}$, for which the anti-symmetric upper-triangular matrix elements ($j<k$) are given by:
\begin{align*}
      &A_{jk}\\
     =&\begin{cases}
        \bra{\psi_{ins}(t)}\beta_j\beta_k\ket{\psi_{ins}(t)} & j \leq 2N, k \leq 2N\\
        \bra{\psi_{ins}(t)}\beta_j\beta^\dagger_{4N+1-k}\ket{\psi_{ins}(t)} & j \leq 2N, k > 2N\\
        \bra{\psi_{ins}(t)}\beta^\dagger_{4N+1-j}\beta^\dagger_{4N+1-k}\ket{\psi_{ins}(t)} & j > 2N, k > 2N
    \end{cases}
\end{align*}
To find elements of $A$ we need to find the correlations, we have $\tilde b(t) = B(t) \tilde C$ and $\tilde \beta(t) = \bar B(t) \tilde C$ from which we can write $\tilde \beta(t) = G(t) \tilde b(t)$ with this we can find out the elements as:
\begin{align*}
    &\bra{\psi_{ins}(t)}\tilde \beta_j\tilde \beta_k\ket{\psi_{ins}(t)} \\
    = &\sum_{m,n}\bra{\psi_{ins}(t)}G_{jm}\tilde b_m G_{kn}\tilde b_n\ket{\psi_{ins}(t)} \\
    = &\sum_{m,n}G_{jm}D_{mn}G_{kn}
\end{align*}
where the matrix $D$ is the correlation matrix of instantaneous creation and annihilation operators on instantaneous ground state and the elements are given by:

\begin{align*}
    D_{mn} &= \bra{\psi_{ins}(t)} \tilde{b}_m \tilde{b}_n \ket{\psi_{ins}(t)} \\
    &= \begin{cases}
        \bra{\psi_{ins}(t)} b_m b_n \ket{\psi_{ins}(t)} = 0, \\
        \qquad \text{if } m \leq 2N \text{ and } n \leq 2N, \\
        \bra{\psi_{ins}(t)} b_m b^\dagger_{4N+1-n} \ket{\psi_{ins}(t)} = \delta_{m,4N+1-n}, \\
        \qquad \text{if } m \leq 2N \text{ and } n > 2N, \\
        \bra{\psi_{ins}(t)} b^\dagger_{4N+1-m} b_n \ket{\psi_{ins}(t)} = 0, \\
        \qquad \text{if } m > 2N \text{ and } n \leq 2N, \\
        \bra{\psi_{ins}(t)} b^\dagger_{4N+1-m} b^\dagger_{4N+1-n} \ket{\psi_{ins}(t)} = 0, \\
        \qquad \text{if } m > 2N \text{ and } n > 2N.
    \end{cases}
\end{align*}

In the manuscript along with the square overlap of time evolved ground state with instantaneous ground state ($\psi_{ins}(t)$), we also calculate the square overlap of time evolved ground state with different parity sector ($\psi'_{ins}(t)$) of the ground state which is the first excited state as the majorana energy splitting is non zero due to the finite size of the system. The correlation matrix for that state has only two elements changed due to the excitation of BdG quasi-particle $\bra{\psi'_{ins}(t)} b_{2N} b^\dagger_{2N}\ket{\psi'_{ins}(t)} = 0$ and $\bra{\psi'_{ins}(t)} b^\dagger_{2N} b_{2N}\ket{\psi'_{ins}(t)} = 1$.
\section{Noise protocol\label{app:noise}}
We considered a dichotomous noise, denoted as $\eta(t)$, oscillates between two values, $a$ and $b$, with switching rates $\mu_a$ and $\mu_b$ respectively. The master equation describing this noise is given by:
\begin{equation}
\frac{\partial P(a, t|x, t_0)}{\partial t} = -\mu_a P(a, t|x, t_0) + \mu_b P(b, t|x, t_0),
\end{equation}
where the solution can be determined by the total probability condition $P(a, t|x, t_0) + P(b, t|x, t_0) = 1$, and the initial condition, $P(x',t|x, t_0) = \delta_{x' x}$. The solution to this master equation is given by:
\begin{align}
P(a, t|x, t_0) &= \frac{\mu_b}{\mu_a + \mu_b} + \left(\frac{\mu_a}{\mu_a + \mu_b} \delta_{ax} - \frac{\mu_b}{\mu_b + \mu_a} \delta_{bx} \right) \nonumber\\
&\quad \times \exp(-(\mu_a + \mu_b)(t - t_0)).
\end{align}
and
\begin{align}
P(b, t|x, t_0) &= \frac{\mu_a}{\mu_a + \mu_b} - \left(\frac{\mu_a}{\mu_a + \mu_b} \delta_{ax} - \frac{\mu_b}{\mu_b + \mu_a} \delta_{bx} \right) \nonumber\\
&\quad \times \exp(-(\mu_a + \mu_b)(t - t_0)).
\end{align}
The stationary solutions of the master equations are straightforwardly given by $P_s(a) = \mu_b/(\mu_a + \mu_b)$ and $P_s(b) =\mu_a/(\mu_a + \mu_b)$.
Subsequently, the stationary mean can be calculated as:
\begin{equation}
\langle \eta(t) \rangle_s = \frac{a\mu_b + b\mu_a}{\mu_a + \mu_b},
\end{equation}
and the stationary time correlation function is:
\begin{align}
\langle \eta(t) \eta(t') \rangle &= \left(\frac{b\mu_a + a\mu_b}{\mu_b + \mu_a}\right)^2 \nonumber\\
&\quad + \frac{\mu_a \mu_b (a - b)^2}{(\mu_a + \mu_b)^2} \exp(-(\mu_a + \mu_b) |t - t'|).
\end{align}
We want the mean and correlation function of the dichotomous noise should satisfy the conditions:
\begin{equation}
\langle \eta(t) \rangle = 0,
\end{equation}
\begin{equation}
\langle \eta(t) \eta(t') \rangle = \sigma^2 \exp\left(-\frac{|\, t - t' \,|}{\tau_c}\right),
\end{equation}
where $\sigma^2$ is the intensity of the noise and $\tau_c$ is the characteristic time. From above equations it is clear that we must have 
$\tau_c = 1/(\mu_a + \mu_b)$. Furthermore, the condition for the mean to be zero is $a\mu_b + b\mu_a = 0$.
The intensity $\sigma^2$ of the noise is given as:
\begin{equation}
\sigma^2 = \frac{(a - b)^2 \mu_a \mu_b}{(\mu_a + \mu_b)^2}.
\end{equation}
Taking these conditions into account, to define a noise, we need to specify three independent variables: $a$, $b$, and $\tau_c$. The conditional probabilities $P_{aa}$ and $P_{ba}$ can be expressed as follows:
\begin{align}
P_{aa} &= P(a, t_{n+1} | a, t_n)\nonumber\\
&= \frac{\mu_b}{\mu_b + \mu_a} + \frac{\mu_a}{\mu_b + \mu_a} \exp(-(\mu_a + \mu_b) \Delta t), \\
P_{bb} &= P(a, t_{n+1} | b, t_n)\nonumber\\
&= \frac{\mu_a}{\mu_b + \mu_a} + \frac{\mu_b}{\mu_b + \mu_a} \exp(-(\mu_a + \mu_b) \Delta t),
\end{align}
where $t_{n+1} = t_n + \Delta t$. The algorithm used to generate the sequence $\eta(t)$ is described as follows: Initially, the starting position of the dichotomous noise is assumed to be $x_0 = a$. Next, a random number $R_n$ distributed uniformly between the interval $[0, 1]$ is generated in the computer. This random number compares with the conditional probabilities $P_{aa}$ or $P_{bb}$. The algorithm then explicitly checks whether the last state $x_n = a$ (for $n = 1, 2, \ldots$). If this condition is true, the algorithm compares $R_n$ with $P_{aa}$. If $R_n < P_{aa}$, then $x_{n+1} = a$; otherwise, $x_{n+1} = b$. Conversely, when the last state $x_n = b$, the algorithm compares $R_n$ with $P_{bb}$. If $R_n < P_{bb}$, then $x_{n+1} = b$; otherwise, $x_{n+1} = a$. The time step for the two next points of the dichotomous noise is $\Delta t$, which is much smaller.

\bibliography{Moving1.bib}

\begin{thebibliography}{58}%
\makeatletter
\providecommand \@ifxundefined [1]{%
 \@ifx{#1\undefined}
}%
\providecommand \@ifnum [1]{%
 \ifnum #1\expandafter \@firstoftwo
 \else \expandafter \@secondoftwo
 \fi
}%
\providecommand \@ifx [1]{%
 \ifx #1\expandafter \@firstoftwo
 \else \expandafter \@secondoftwo
 \fi
}%
\providecommand \natexlab [1]{#1}%
\providecommand \enquote  [1]{``#1''}%
\providecommand \bibnamefont  [1]{#1}%
\providecommand \bibfnamefont [1]{#1}%
\providecommand \citenamefont [1]{#1}%
\providecommand \href@noop [0]{\@secondoftwo}%
\providecommand \href [0]{\begingroup \@sanitize@url \@href}%
\providecommand \@href[1]{\@@startlink{#1}\@@href}%
\providecommand \@@href[1]{\endgroup#1\@@endlink}%
\providecommand \@sanitize@url [0]{\catcode `\\12\catcode `\$12\catcode
  `\&12\catcode `\#12\catcode `\^12\catcode `\_12\catcode `\%12\relax}%
\providecommand \@@startlink[1]{}%
\providecommand \@@endlink[0]{}%
\providecommand \url  [0]{\begingroup\@sanitize@url \@url }%
\providecommand \@url [1]{\endgroup\@href {#1}{\urlprefix }}%
\providecommand \urlprefix  [0]{URL }%
\providecommand \Eprint [0]{\href }%
\providecommand \doibase [0]{https://doi.org/}%
\providecommand \selectlanguage [0]{\@gobble}%
\providecommand \bibinfo  [0]{\@secondoftwo}%
\providecommand \bibfield  [0]{\@secondoftwo}%
\providecommand \translation [1]{[#1]}%
\providecommand \BibitemOpen [0]{}%
\providecommand \bibitemStop [0]{}%
\providecommand \bibitemNoStop [0]{.\EOS\space}%
\providecommand \EOS [0]{\spacefactor3000\relax}%
\providecommand \BibitemShut  [1]{\csname bibitem#1\endcsname}%
\let\auto@bib@innerbib\@empty
\bibitem [{\citenamefont {Kitaev}(2001)}]{kitaev_unpaired_2001}%
  \BibitemOpen
  \bibfield  {author} {\bibinfo {author} {\bibfnamefont {A.~Y.}\ \bibnamefont
  {Kitaev}},\ }\bibfield  {title} {\bibinfo {title} {Unpaired {Majorana}
  fermions in quantum wires},\ }\href@noop {} {\bibfield  {journal} {\bibinfo
  {journal} {Physics-uspekhi}\ }\textbf {\bibinfo {volume} {44}},\ \bibinfo
  {pages} {131} (\bibinfo {year} {2001})},\ \bibinfo {note} {publisher: IOP
  Publishing}\BibitemShut {NoStop}%
\bibitem [{\citenamefont {Nayak}\ \emph {et~al.}(2008)\citenamefont {Nayak},
  \citenamefont {Simon}, \citenamefont {Stern}, \citenamefont {Freedman},\ and\
  \citenamefont {Das~Sarma}}]{nayak_non-abelian_2008}%
  \BibitemOpen
  \bibfield  {author} {\bibinfo {author} {\bibfnamefont {C.}~\bibnamefont
  {Nayak}}, \bibinfo {author} {\bibfnamefont {S.~H.}\ \bibnamefont {Simon}},
  \bibinfo {author} {\bibfnamefont {A.}~\bibnamefont {Stern}}, \bibinfo
  {author} {\bibfnamefont {M.}~\bibnamefont {Freedman}},\ and\ \bibinfo
  {author} {\bibfnamefont {S.}~\bibnamefont {Das~Sarma}},\ }\bibfield  {title}
  {\bibinfo {title} {Non-{Abelian} anyons and topological quantum
  computation},\ }\href {https://doi.org/10.1103/RevModPhys.80.1083} {\bibfield
   {journal} {\bibinfo  {journal} {Rev. Mod. Phys.}\ }\textbf {\bibinfo
  {volume} {80}},\ \bibinfo {pages} {1083} (\bibinfo {year} {2008})},\ \bibinfo
  {note} {publisher: American Physical Society}\BibitemShut {NoStop}%
\bibitem [{\citenamefont {Alicea}(2012)}]{alicea_new_2012}%
  \BibitemOpen
  \bibfield  {author} {\bibinfo {author} {\bibfnamefont {J.}~\bibnamefont
  {Alicea}},\ }\bibfield  {title} {\bibinfo {title} {New directions in the
  pursuit of {Majorana} fermions in solid state systems},\ }\href@noop {}
  {\bibfield  {journal} {\bibinfo  {journal} {Reports on progress in physics}\
  }\textbf {\bibinfo {volume} {75}},\ \bibinfo {pages} {076501} (\bibinfo
  {year} {2012})},\ \bibinfo {note} {publisher: IOP Publishing}\BibitemShut
  {NoStop}%
\bibitem [{\citenamefont {Sarma}\ \emph {et~al.}(2015)\citenamefont {Sarma},
  \citenamefont {Freedman},\ and\ \citenamefont {Nayak}}]{sarma_majorana_2015}%
  \BibitemOpen
  \bibfield  {author} {\bibinfo {author} {\bibfnamefont {S.~D.}\ \bibnamefont
  {Sarma}}, \bibinfo {author} {\bibfnamefont {M.}~\bibnamefont {Freedman}},\
  and\ \bibinfo {author} {\bibfnamefont {C.}~\bibnamefont {Nayak}},\ }\bibfield
   {title} {\bibinfo {title} {Majorana zero modes and topological quantum
  computation},\ }\href@noop {} {\bibfield  {journal} {\bibinfo  {journal} {npj
  Quantum Information}\ }\textbf {\bibinfo {volume} {1}},\ \bibinfo {pages} {1}
  (\bibinfo {year} {2015})},\ \bibinfo {note} {publisher: Nature Publishing
  Group}\BibitemShut {NoStop}%
\bibitem [{\citenamefont {Prada}\ \emph {et~al.}(2020)\citenamefont {Prada},
  \citenamefont {San-Jose}, \citenamefont {de~Moor}, \citenamefont {Geresdi},
  \citenamefont {Lee}, \citenamefont {Klinovaja}, \citenamefont {Loss},
  \citenamefont {Nygård}, \citenamefont {Aguado},\ and\ \citenamefont
  {Kouwenhoven}}]{prada_andreev_2020}%
  \BibitemOpen
  \bibfield  {author} {\bibinfo {author} {\bibfnamefont {E.}~\bibnamefont
  {Prada}}, \bibinfo {author} {\bibfnamefont {P.}~\bibnamefont {San-Jose}},
  \bibinfo {author} {\bibfnamefont {M.~W.}\ \bibnamefont {de~Moor}}, \bibinfo
  {author} {\bibfnamefont {A.}~\bibnamefont {Geresdi}}, \bibinfo {author}
  {\bibfnamefont {E.~J.}\ \bibnamefont {Lee}}, \bibinfo {author} {\bibfnamefont
  {J.}~\bibnamefont {Klinovaja}}, \bibinfo {author} {\bibfnamefont
  {D.}~\bibnamefont {Loss}}, \bibinfo {author} {\bibfnamefont {J.}~\bibnamefont
  {Nygård}}, \bibinfo {author} {\bibfnamefont {R.}~\bibnamefont {Aguado}},\
  and\ \bibinfo {author} {\bibfnamefont {L.~P.}\ \bibnamefont {Kouwenhoven}},\
  }\bibfield  {title} {\bibinfo {title} {From {Andreev} to {Majorana} bound
  states in hybrid superconductor–semiconductor nanowires},\ }\href@noop {}
  {\bibfield  {journal} {\bibinfo  {journal} {Nature Reviews Physics}\ }\textbf
  {\bibinfo {volume} {2}},\ \bibinfo {pages} {575} (\bibinfo {year} {2020})},\
  \bibinfo {note} {publisher: Nature Publishing Group}\BibitemShut {NoStop}%
\bibitem [{\citenamefont {Moore}\ and\ \citenamefont
  {Read}(1991)}]{moore_nonabelions_1991}%
  \BibitemOpen
  \bibfield  {author} {\bibinfo {author} {\bibfnamefont {G.}~\bibnamefont
  {Moore}}\ and\ \bibinfo {author} {\bibfnamefont {N.}~\bibnamefont {Read}},\
  }\bibfield  {title} {\bibinfo {title} {Nonabelions in the fractional quantum
  hall effect},\ }\href
  {https://doi.org/https://doi.org/10.1016/0550-3213(91)90407-O} {\bibfield
  {journal} {\bibinfo  {journal} {Nuclear Physics B}\ }\textbf {\bibinfo
  {volume} {360}},\ \bibinfo {pages} {362} (\bibinfo {year}
  {1991})}\BibitemShut {NoStop}%
\bibitem [{\citenamefont {Rice}\ and\ \citenamefont
  {Sigrist}(1995)}]{rice_sr2ruo4_1995}%
  \BibitemOpen
  \bibfield  {author} {\bibinfo {author} {\bibfnamefont {T.}~\bibnamefont
  {Rice}}\ and\ \bibinfo {author} {\bibfnamefont {M.}~\bibnamefont {Sigrist}},\
  }\bibfield  {title} {\bibinfo {title} {{Sr2RuO4}: an electronic analogue of
  {3He}?},\ }\href@noop {} {\bibfield  {journal} {\bibinfo  {journal} {Journal
  of Physics: Condensed Matter}\ }\textbf {\bibinfo {volume} {7}},\ \bibinfo
  {pages} {L643} (\bibinfo {year} {1995})},\ \bibinfo {note} {publisher: IOP
  Publishing}\BibitemShut {NoStop}%
\bibitem [{\citenamefont {Fu}\ and\ \citenamefont
  {Kane}(2008)}]{fu_superconducting_2008}%
  \BibitemOpen
  \bibfield  {author} {\bibinfo {author} {\bibfnamefont {L.}~\bibnamefont
  {Fu}}\ and\ \bibinfo {author} {\bibfnamefont {C.~L.}\ \bibnamefont {Kane}},\
  }\bibfield  {title} {\bibinfo {title} {Superconducting {Proximity} {Effect}
  and {Majorana} {Fermions} at the {Surface} of a {Topological} {Insulator}},\
  }\href {https://doi.org/10.1103/PhysRevLett.100.096407} {\bibfield  {journal}
  {\bibinfo  {journal} {Phys. Rev. Lett.}\ }\textbf {\bibinfo {volume} {100}},\
  \bibinfo {pages} {096407} (\bibinfo {year} {2008})},\ \bibinfo {note}
  {publisher: American Physical Society}\BibitemShut {NoStop}%
\bibitem [{\citenamefont {Sau}\ \emph {et~al.}(2010)\citenamefont {Sau},
  \citenamefont {Lutchyn}, \citenamefont {Tewari},\ and\ \citenamefont
  {Sarma}}]{sau_generic_2010}%
  \BibitemOpen
  \bibfield  {author} {\bibinfo {author} {\bibfnamefont {J.~D.}\ \bibnamefont
  {Sau}}, \bibinfo {author} {\bibfnamefont {R.~M.}\ \bibnamefont {Lutchyn}},
  \bibinfo {author} {\bibfnamefont {S.}~\bibnamefont {Tewari}},\ and\ \bibinfo
  {author} {\bibfnamefont {S.~D.}\ \bibnamefont {Sarma}},\ }\bibfield  {title}
  {\bibinfo {title} {Generic new platform for topological quantum computation
  using semiconductor heterostructures},\ }\bibfield  {journal} {\bibinfo
  {journal} {Physical Review Letters}\ }\textbf {\bibinfo {volume} {104}},\
  \href {https://doi.org/10.1103/PHYSREVLETT.104.040502}
  {10.1103/PHYSREVLETT.104.040502} (\bibinfo {year} {2010})\BibitemShut
  {NoStop}%
\bibitem [{\citenamefont {Oreg}\ \emph {et~al.}(2010)\citenamefont {Oreg},
  \citenamefont {Refael},\ and\ \citenamefont {von Oppen}}]{oreg_helical_2010}%
  \BibitemOpen
  \bibfield  {author} {\bibinfo {author} {\bibfnamefont {Y.}~\bibnamefont
  {Oreg}}, \bibinfo {author} {\bibfnamefont {G.}~\bibnamefont {Refael}},\ and\
  \bibinfo {author} {\bibfnamefont {F.}~\bibnamefont {von Oppen}},\ }\bibfield
  {title} {\bibinfo {title} {Helical {Liquids} and {Majorana} {Bound} {States}
  in {Quantum} {Wires}},\ }\href
  {https://doi.org/10.1103/PhysRevLett.105.177002} {\bibfield  {journal}
  {\bibinfo  {journal} {Phys. Rev. Lett.}\ }\textbf {\bibinfo {volume} {105}},\
  \bibinfo {pages} {177002} (\bibinfo {year} {2010})},\ \bibinfo {note}
  {publisher: American Physical Society}\BibitemShut {NoStop}%
\bibitem [{\citenamefont {Stanescu}\ \emph {et~al.}(2010)\citenamefont
  {Stanescu}, \citenamefont {Sau}, \citenamefont {Lutchyn},\ and\ \citenamefont
  {Sarma}}]{stanescu_proximity_2010}%
  \BibitemOpen
  \bibfield  {author} {\bibinfo {author} {\bibfnamefont {T.~D.}\ \bibnamefont
  {Stanescu}}, \bibinfo {author} {\bibfnamefont {J.~D.}\ \bibnamefont {Sau}},
  \bibinfo {author} {\bibfnamefont {R.~M.}\ \bibnamefont {Lutchyn}},\ and\
  \bibinfo {author} {\bibfnamefont {S.~D.}\ \bibnamefont {Sarma}},\ }\bibfield
  {title} {\bibinfo {title} {Proximity effect at the superconductor-topological
  insulator interface},\ }\bibfield  {journal} {\bibinfo  {journal} {Physical
  Review B - Condensed Matter and Materials Physics}\ }\textbf {\bibinfo
  {volume} {81}},\ \href {https://doi.org/10.1103/PHYSREVB.81.241310}
  {10.1103/PHYSREVB.81.241310} (\bibinfo {year} {2010})\BibitemShut {NoStop}%
\bibitem [{\citenamefont {Lutchyn}\ \emph {et~al.}(2010)\citenamefont
  {Lutchyn}, \citenamefont {Sau},\ and\ \citenamefont
  {Das~Sarma}}]{lutchyn_majorana_2010}%
  \BibitemOpen
  \bibfield  {author} {\bibinfo {author} {\bibfnamefont {R.~M.}\ \bibnamefont
  {Lutchyn}}, \bibinfo {author} {\bibfnamefont {J.~D.}\ \bibnamefont {Sau}},\
  and\ \bibinfo {author} {\bibfnamefont {S.}~\bibnamefont {Das~Sarma}},\
  }\bibfield  {title} {\bibinfo {title} {Majorana {Fermions} and a
  {Topological} {Phase} {Transition} in {Semiconductor}-{Superconductor}
  {Heterostructures}},\ }\href {https://doi.org/10.1103/PhysRevLett.105.077001}
  {\bibfield  {journal} {\bibinfo  {journal} {Phys. Rev. Lett.}\ }\textbf
  {\bibinfo {volume} {105}},\ \bibinfo {pages} {077001} (\bibinfo {year}
  {2010})},\ \bibinfo {note} {publisher: American Physical Society}\BibitemShut
  {NoStop}%
\bibitem [{\citenamefont {Dmytruk}\ \emph {et~al.}(2018)\citenamefont
  {Dmytruk}, \citenamefont {Chevallier}, \citenamefont {Loss},\ and\
  \citenamefont {Klinovaja}}]{dmytruk_renormalization_2018}%
  \BibitemOpen
  \bibfield  {author} {\bibinfo {author} {\bibfnamefont {O.}~\bibnamefont
  {Dmytruk}}, \bibinfo {author} {\bibfnamefont {D.}~\bibnamefont {Chevallier}},
  \bibinfo {author} {\bibfnamefont {D.}~\bibnamefont {Loss}},\ and\ \bibinfo
  {author} {\bibfnamefont {J.}~\bibnamefont {Klinovaja}},\ }\bibfield  {title}
  {\bibinfo {title} {Renormalization of the quantum dot g -factor in
  superconducting {Rashba} nanowires},\ }\href
  {https://doi.org/10.1103/PhysRevB.98.165403} {\bibfield  {journal} {\bibinfo
  {journal} {Physical Review B}\ }\textbf {\bibinfo {volume} {98}},\ \bibinfo
  {pages} {165403} (\bibinfo {year} {2018})}\BibitemShut {NoStop}%
\bibitem [{\citenamefont {Albrecht}\ \emph {et~al.}(2016)\citenamefont
  {Albrecht}, \citenamefont {Higginbotham}, \citenamefont {Madsen},
  \citenamefont {Kuemmeth}, \citenamefont {Jespersen}, \citenamefont {Nygård},
  \citenamefont {Krogstrup},\ and\ \citenamefont
  {Marcus}}]{albrecht_exponential_2016}%
  \BibitemOpen
  \bibfield  {author} {\bibinfo {author} {\bibfnamefont {S.~M.}\ \bibnamefont
  {Albrecht}}, \bibinfo {author} {\bibfnamefont {A.~P.}\ \bibnamefont
  {Higginbotham}}, \bibinfo {author} {\bibfnamefont {M.}~\bibnamefont
  {Madsen}}, \bibinfo {author} {\bibfnamefont {F.}~\bibnamefont {Kuemmeth}},
  \bibinfo {author} {\bibfnamefont {T.~S.}\ \bibnamefont {Jespersen}}, \bibinfo
  {author} {\bibfnamefont {J.}~\bibnamefont {Nygård}}, \bibinfo {author}
  {\bibfnamefont {P.}~\bibnamefont {Krogstrup}},\ and\ \bibinfo {author}
  {\bibfnamefont {C.}~\bibnamefont {Marcus}},\ }\bibfield  {title} {\bibinfo
  {title} {Exponential protection of zero modes in {Majorana} islands},\
  }\href@noop {} {\bibfield  {journal} {\bibinfo  {journal} {Nature}\ }\textbf
  {\bibinfo {volume} {531}},\ \bibinfo {pages} {206} (\bibinfo {year}
  {2016})},\ \bibinfo {note} {publisher: Nature Publishing Group}\BibitemShut
  {NoStop}%
\bibitem [{\citenamefont {Churchill}\ \emph {et~al.}(2013)\citenamefont
  {Churchill}, \citenamefont {Fatemi}, \citenamefont {Grove-Rasmussen},
  \citenamefont {Deng}, \citenamefont {Caroff}, \citenamefont {Xu},\ and\
  \citenamefont {Marcus}}]{churchill_superconductor-nanowire_2013}%
  \BibitemOpen
  \bibfield  {author} {\bibinfo {author} {\bibfnamefont {H.~O.~H.}\
  \bibnamefont {Churchill}}, \bibinfo {author} {\bibfnamefont {V.}~\bibnamefont
  {Fatemi}}, \bibinfo {author} {\bibfnamefont {K.}~\bibnamefont
  {Grove-Rasmussen}}, \bibinfo {author} {\bibfnamefont {M.~T.}\ \bibnamefont
  {Deng}}, \bibinfo {author} {\bibfnamefont {P.}~\bibnamefont {Caroff}},
  \bibinfo {author} {\bibfnamefont {H.~Q.}\ \bibnamefont {Xu}},\ and\ \bibinfo
  {author} {\bibfnamefont {C.~M.}\ \bibnamefont {Marcus}},\ }\bibfield  {title}
  {\bibinfo {title} {Superconductor-nanowire devices from tunneling to the
  multichannel regime: {Zero}-bias oscillations and magnetoconductance
  crossover},\ }\href {https://doi.org/10.1103/PhysRevB.87.241401} {\bibfield
  {journal} {\bibinfo  {journal} {Phys. Rev. B}\ }\textbf {\bibinfo {volume}
  {87}},\ \bibinfo {pages} {241401} (\bibinfo {year} {2013})},\ \bibinfo {note}
  {publisher: American Physical Society}\BibitemShut {NoStop}%
\bibitem [{\citenamefont {Das}\ \emph {et~al.}(2012)\citenamefont {Das},
  \citenamefont {Ronen}, \citenamefont {Most}, \citenamefont {Oreg},
  \citenamefont {Heiblum},\ and\ \citenamefont
  {Shtrikman}}]{das_zero-bias_2012}%
  \BibitemOpen
  \bibfield  {author} {\bibinfo {author} {\bibfnamefont {A.}~\bibnamefont
  {Das}}, \bibinfo {author} {\bibfnamefont {Y.}~\bibnamefont {Ronen}}, \bibinfo
  {author} {\bibfnamefont {Y.}~\bibnamefont {Most}}, \bibinfo {author}
  {\bibfnamefont {Y.}~\bibnamefont {Oreg}}, \bibinfo {author} {\bibfnamefont
  {M.}~\bibnamefont {Heiblum}},\ and\ \bibinfo {author} {\bibfnamefont
  {H.}~\bibnamefont {Shtrikman}},\ }\bibfield  {title} {\bibinfo {title}
  {Zero-bias peaks and splitting in an {Al}–{InAs} nanowire topological
  superconductor as a signature of {Majorana} fermions},\ }\href@noop {}
  {\bibfield  {journal} {\bibinfo  {journal} {Nature Physics}\ }\textbf
  {\bibinfo {volume} {8}},\ \bibinfo {pages} {887} (\bibinfo {year} {2012})},\
  \bibinfo {note} {publisher: Nature Publishing Group}\BibitemShut {NoStop}%
\bibitem [{\citenamefont {Mourik}\ \emph {et~al.}(2012)\citenamefont {Mourik},
  \citenamefont {Zuo}, \citenamefont {Frolov}, \citenamefont {Plissard},
  \citenamefont {Bakkers},\ and\ \citenamefont
  {Kouwenhoven}}]{mourik_signatures_2012}%
  \BibitemOpen
  \bibfield  {author} {\bibinfo {author} {\bibfnamefont {V.}~\bibnamefont
  {Mourik}}, \bibinfo {author} {\bibfnamefont {K.}~\bibnamefont {Zuo}},
  \bibinfo {author} {\bibfnamefont {S.~M.}\ \bibnamefont {Frolov}}, \bibinfo
  {author} {\bibfnamefont {S.}~\bibnamefont {Plissard}}, \bibinfo {author}
  {\bibfnamefont {E.~P.}\ \bibnamefont {Bakkers}},\ and\ \bibinfo {author}
  {\bibfnamefont {L.~P.}\ \bibnamefont {Kouwenhoven}},\ }\bibfield  {title}
  {\bibinfo {title} {Signatures of {Majorana} fermions in hybrid
  superconductor-semiconductor nanowire devices},\ }\href@noop {} {\bibfield
  {journal} {\bibinfo  {journal} {Science}\ }\textbf {\bibinfo {volume}
  {336}},\ \bibinfo {pages} {1003} (\bibinfo {year} {2012})},\ \bibinfo {note}
  {publisher: American Association for the Advancement of Science}\BibitemShut
  {NoStop}%
\bibitem [{\citenamefont {Reeg}\ \emph {et~al.}(2018)\citenamefont {Reeg},
  \citenamefont {Dmytruk}, \citenamefont {Chevallier}, \citenamefont {Loss},\
  and\ \citenamefont {Klinovaja}}]{reeg_zero-energy_2018}%
  \BibitemOpen
  \bibfield  {author} {\bibinfo {author} {\bibfnamefont {C.}~\bibnamefont
  {Reeg}}, \bibinfo {author} {\bibfnamefont {O.}~\bibnamefont {Dmytruk}},
  \bibinfo {author} {\bibfnamefont {D.}~\bibnamefont {Chevallier}}, \bibinfo
  {author} {\bibfnamefont {D.}~\bibnamefont {Loss}},\ and\ \bibinfo {author}
  {\bibfnamefont {J.}~\bibnamefont {Klinovaja}},\ }\bibfield  {title} {\bibinfo
  {title} {Zero-energy {Andreev} bound states from quantum dots in proximitized
  {Rashba} nanowires},\ }\href {https://doi.org/10.1103/PhysRevB.98.245407}
  {\bibfield  {journal} {\bibinfo  {journal} {Phys. Rev. B}\ }\textbf {\bibinfo
  {volume} {98}},\ \bibinfo {pages} {245407} (\bibinfo {year} {2018})},\
  \bibinfo {note} {publisher: American Physical Society}\BibitemShut {NoStop}%
\bibitem [{\citenamefont {Dmytruk}\ \emph {et~al.}(2020)\citenamefont
  {Dmytruk}, \citenamefont {Loss},\ and\ \citenamefont
  {Klinovaja}}]{dmytruk_pinning_2020}%
  \BibitemOpen
  \bibfield  {author} {\bibinfo {author} {\bibfnamefont {O.}~\bibnamefont
  {Dmytruk}}, \bibinfo {author} {\bibfnamefont {D.}~\bibnamefont {Loss}},\ and\
  \bibinfo {author} {\bibfnamefont {J.}~\bibnamefont {Klinovaja}},\ }\bibfield
  {title} {\bibinfo {title} {Pinning of {Andreev} bound states to zero energy
  in two-dimensional superconductor- semiconductor {Rashba} heterostructures},\
  }\href {https://doi.org/10.1103/PhysRevB.102.245431} {\bibfield  {journal}
  {\bibinfo  {journal} {Phys. Rev. B}\ }\textbf {\bibinfo {volume} {102}},\
  \bibinfo {pages} {245431} (\bibinfo {year} {2020})},\ \bibinfo {note}
  {publisher: American Physical Society}\BibitemShut {NoStop}%
\bibitem [{\citenamefont {Sahu}\ \emph {et~al.}(2023)\citenamefont {Sahu},
  \citenamefont {Khade},\ and\ \citenamefont
  {Gangadharaiah}}]{sahu_effect_2023}%
  \BibitemOpen
  \bibfield  {author} {\bibinfo {author} {\bibfnamefont {D.}~\bibnamefont
  {Sahu}}, \bibinfo {author} {\bibfnamefont {V.}~\bibnamefont {Khade}},\ and\
  \bibinfo {author} {\bibfnamefont {S.}~\bibnamefont {Gangadharaiah}},\
  }\bibfield  {title} {\bibinfo {title} {Effect of topological length on bound
  state signatures in a topological nanowire},\ }\href
  {https://doi.org/10.1103/PhysRevB.108.205426} {\bibfield  {journal} {\bibinfo
   {journal} {Physical Review B}\ }\textbf {\bibinfo {volume} {108}},\ \bibinfo
  {pages} {205426} (\bibinfo {year} {2023})},\ \bibinfo {note} {publisher:
  American Physical Society}\BibitemShut {NoStop}%
\bibitem [{\citenamefont {Alicea}\ \emph {et~al.}(2011)\citenamefont {Alicea},
  \citenamefont {Oreg}, \citenamefont {Refael}, \citenamefont {von Oppen},\
  and\ \citenamefont {Fisher}}]{alicea_non-abelian_2011}%
  \BibitemOpen
  \bibfield  {author} {\bibinfo {author} {\bibfnamefont {J.}~\bibnamefont
  {Alicea}}, \bibinfo {author} {\bibfnamefont {Y.}~\bibnamefont {Oreg}},
  \bibinfo {author} {\bibfnamefont {G.}~\bibnamefont {Refael}}, \bibinfo
  {author} {\bibfnamefont {F.}~\bibnamefont {von Oppen}},\ and\ \bibinfo
  {author} {\bibfnamefont {M.~P.~A.}\ \bibnamefont {Fisher}},\ }\bibfield
  {title} {\bibinfo {title} {Non-{Abelian} statistics and topological quantum
  information processing in {1D} wire networks},\ }\href
  {https://doi.org/10.1038/nphys1915} {\bibfield  {journal} {\bibinfo
  {journal} {Nature Physics}\ }\textbf {\bibinfo {volume} {7}},\ \bibinfo
  {pages} {412} (\bibinfo {year} {2011})},\ \bibinfo {note} {publisher: Nature
  Publishing Group}\BibitemShut {NoStop}%
\bibitem [{\citenamefont {Sau}\ \emph {et~al.}(2011)\citenamefont {Sau},
  \citenamefont {Clarke},\ and\ \citenamefont {Tewari}}]{sau_controlling_2011}%
  \BibitemOpen
  \bibfield  {author} {\bibinfo {author} {\bibfnamefont {J.~D.}\ \bibnamefont
  {Sau}}, \bibinfo {author} {\bibfnamefont {D.~J.}\ \bibnamefont {Clarke}},\
  and\ \bibinfo {author} {\bibfnamefont {S.}~\bibnamefont {Tewari}},\
  }\bibfield  {title} {\bibinfo {title} {Controlling non-{Abelian} statistics
  of {Majorana} fermions in semiconductor nanowires},\ }\href
  {https://doi.org/10.1103/PhysRevB.84.094505} {\bibfield  {journal} {\bibinfo
  {journal} {Physical Review B}\ }\textbf {\bibinfo {volume} {84}},\ \bibinfo
  {pages} {094505} (\bibinfo {year} {2011})},\ \bibinfo {note} {publisher:
  American Physical Society}\BibitemShut {NoStop}%
\bibitem [{\citenamefont {Halperin}\ \emph {et~al.}(2012)\citenamefont
  {Halperin}, \citenamefont {Oreg}, \citenamefont {Stern}, \citenamefont
  {Refael}, \citenamefont {Alicea},\ and\ \citenamefont {von
  Oppen}}]{halperin_adiabatic_2012}%
  \BibitemOpen
  \bibfield  {author} {\bibinfo {author} {\bibfnamefont {B.~I.}\ \bibnamefont
  {Halperin}}, \bibinfo {author} {\bibfnamefont {Y.}~\bibnamefont {Oreg}},
  \bibinfo {author} {\bibfnamefont {A.}~\bibnamefont {Stern}}, \bibinfo
  {author} {\bibfnamefont {G.}~\bibnamefont {Refael}}, \bibinfo {author}
  {\bibfnamefont {J.}~\bibnamefont {Alicea}},\ and\ \bibinfo {author}
  {\bibfnamefont {F.}~\bibnamefont {von Oppen}},\ }\bibfield  {title} {\bibinfo
  {title} {Adiabatic manipulations of {Majorana} fermions in a
  three-dimensional network of quantum wires},\ }\href
  {https://doi.org/10.1103/PhysRevB.85.144501} {\bibfield  {journal} {\bibinfo
  {journal} {Physical Review B}\ }\textbf {\bibinfo {volume} {85}},\ \bibinfo
  {pages} {144501} (\bibinfo {year} {2012})},\ \bibinfo {note} {publisher:
  American Physical Society}\BibitemShut {NoStop}%
\bibitem [{\citenamefont {Tutschku}\ \emph {et~al.}(2020)\citenamefont
  {Tutschku}, \citenamefont {Reinthaler}, \citenamefont {Lei}, \citenamefont
  {MacDonald},\ and\ \citenamefont
  {Hankiewicz}}]{tutschku_majorana-based_2020}%
  \BibitemOpen
  \bibfield  {author} {\bibinfo {author} {\bibfnamefont {C.}~\bibnamefont
  {Tutschku}}, \bibinfo {author} {\bibfnamefont {R.~W.}\ \bibnamefont
  {Reinthaler}}, \bibinfo {author} {\bibfnamefont {C.}~\bibnamefont {Lei}},
  \bibinfo {author} {\bibfnamefont {A.~H.}\ \bibnamefont {MacDonald}},\ and\
  \bibinfo {author} {\bibfnamefont {E.~M.}\ \bibnamefont {Hankiewicz}},\
  }\bibfield  {title} {\bibinfo {title} {Majorana-based quantum computing in
  nanowire devices},\ }\href {https://doi.org/10.1103/PhysRevB.102.125407}
  {\bibfield  {journal} {\bibinfo  {journal} {Physical Review B}\ }\textbf
  {\bibinfo {volume} {102}},\ \bibinfo {pages} {125407} (\bibinfo {year}
  {2020})}\BibitemShut {NoStop}%
\bibitem [{\citenamefont {Heck}\ \emph {et~al.}(2012)\citenamefont {Heck},
  \citenamefont {Akhmerov}, \citenamefont {Hassler}, \citenamefont {Burrello},\
  and\ \citenamefont {Beenakker}}]{heck_coulomb-assisted_2012}%
  \BibitemOpen
  \bibfield  {author} {\bibinfo {author} {\bibfnamefont {B.~v.}\ \bibnamefont
  {Heck}}, \bibinfo {author} {\bibfnamefont {A.~R.}\ \bibnamefont {Akhmerov}},
  \bibinfo {author} {\bibfnamefont {F.}~\bibnamefont {Hassler}}, \bibinfo
  {author} {\bibfnamefont {M.}~\bibnamefont {Burrello}},\ and\ \bibinfo
  {author} {\bibfnamefont {C.~W.~J.}\ \bibnamefont {Beenakker}},\ }\bibfield
  {title} {\bibinfo {title} {Coulomb-assisted braiding of {Majorana} fermions
  in a {Josephson} junction array},\ }\href
  {https://doi.org/10.1088/1367-2630/14/3/035019} {\bibfield  {journal}
  {\bibinfo  {journal} {New Journal of Physics}\ }\textbf {\bibinfo {volume}
  {14}},\ \bibinfo {pages} {035019} (\bibinfo {year} {2012})},\ \bibinfo {note}
  {publisher: IOP Publishing}\BibitemShut {NoStop}%
\bibitem [{\citenamefont {Hyart}\ \emph {et~al.}(2013)\citenamefont {Hyart},
  \citenamefont {van Heck}, \citenamefont {Fulga}, \citenamefont {Burrello},
  \citenamefont {Akhmerov},\ and\ \citenamefont
  {Beenakker}}]{hyart_flux-controlled_2013}%
  \BibitemOpen
  \bibfield  {author} {\bibinfo {author} {\bibfnamefont {T.}~\bibnamefont
  {Hyart}}, \bibinfo {author} {\bibfnamefont {B.}~\bibnamefont {van Heck}},
  \bibinfo {author} {\bibfnamefont {I.~C.}\ \bibnamefont {Fulga}}, \bibinfo
  {author} {\bibfnamefont {M.}~\bibnamefont {Burrello}}, \bibinfo {author}
  {\bibfnamefont {A.~R.}\ \bibnamefont {Akhmerov}},\ and\ \bibinfo {author}
  {\bibfnamefont {C.~W.~J.}\ \bibnamefont {Beenakker}},\ }\bibfield  {title}
  {\bibinfo {title} {Flux-controlled quantum computation with {Majorana}
  fermions},\ }\href {https://doi.org/10.1103/PhysRevB.88.035121} {\bibfield
  {journal} {\bibinfo  {journal} {Physical Review B}\ }\textbf {\bibinfo
  {volume} {88}},\ \bibinfo {pages} {035121} (\bibinfo {year} {2013})},\
  \bibinfo {note} {publisher: American Physical Society}\BibitemShut {NoStop}%
\bibitem [{\citenamefont {Hegde}\ \emph {et~al.}(2020)\citenamefont {Hegde},
  \citenamefont {Yue}, \citenamefont {Wang}, \citenamefont {Huemiller},
  \citenamefont {Van~Harlingen},\ and\ \citenamefont
  {Vishveshwara}}]{hegde_topological_2020}%
  \BibitemOpen
  \bibfield  {author} {\bibinfo {author} {\bibfnamefont {S.~S.}\ \bibnamefont
  {Hegde}}, \bibinfo {author} {\bibfnamefont {G.}~\bibnamefont {Yue}}, \bibinfo
  {author} {\bibfnamefont {Y.}~\bibnamefont {Wang}}, \bibinfo {author}
  {\bibfnamefont {E.}~\bibnamefont {Huemiller}}, \bibinfo {author}
  {\bibfnamefont {D.~J.}\ \bibnamefont {Van~Harlingen}},\ and\ \bibinfo
  {author} {\bibfnamefont {S.}~\bibnamefont {Vishveshwara}},\ }\bibfield
  {title} {\bibinfo {title} {A topological {Josephson} junction platform for
  creating, manipulating, and braiding {Majorana} bound states},\ }\href
  {https://doi.org/10.1016/j.aop.2020.168326} {\bibfield  {journal} {\bibinfo
  {journal} {Annals of Physics}\ }\textbf {\bibinfo {volume} {423}},\ \bibinfo
  {pages} {168326} (\bibinfo {year} {2020})}\BibitemShut {NoStop}%
\bibitem [{\citenamefont {Boross}\ and\ \citenamefont
  {Pályi}(2024)}]{boross_braiding-based_2024}%
  \BibitemOpen
  \bibfield  {author} {\bibinfo {author} {\bibfnamefont {P.}~\bibnamefont
  {Boross}}\ and\ \bibinfo {author} {\bibfnamefont {A.}~\bibnamefont
  {Pályi}},\ }\bibfield  {title} {\bibinfo {title} {Braiding-based quantum
  control of a {Majorana} qubit built from quantum dots},\ }\href
  {https://doi.org/10.1103/PhysRevB.109.125410} {\bibfield  {journal} {\bibinfo
   {journal} {Physical Review B}\ }\textbf {\bibinfo {volume} {109}},\ \bibinfo
  {pages} {125410} (\bibinfo {year} {2024})},\ \bibinfo {note} {publisher:
  American Physical Society}\BibitemShut {NoStop}%
\bibitem [{\citenamefont {Malciu}\ \emph {et~al.}(2018)\citenamefont {Malciu},
  \citenamefont {Mazza},\ and\ \citenamefont {Mora}}]{malciu_braiding_2018}%
  \BibitemOpen
  \bibfield  {author} {\bibinfo {author} {\bibfnamefont {C.}~\bibnamefont
  {Malciu}}, \bibinfo {author} {\bibfnamefont {L.}~\bibnamefont {Mazza}},\ and\
  \bibinfo {author} {\bibfnamefont {C.}~\bibnamefont {Mora}},\ }\bibfield
  {title} {\bibinfo {title} {Braiding {Majorana} zero modes using quantum
  dots},\ }\href {https://doi.org/10.1103/PhysRevB.98.165426} {\bibfield
  {journal} {\bibinfo  {journal} {Physical Review B}\ }\textbf {\bibinfo
  {volume} {98}},\ \bibinfo {pages} {165426} (\bibinfo {year} {2018})},\
  \bibinfo {note} {publisher: American Physical Society}\BibitemShut {NoStop}%
\bibitem [{\citenamefont {Liu}\ \emph {et~al.}(2017)\citenamefont {Liu},
  \citenamefont {Sau}, \citenamefont {Stanescu},\ and\ \citenamefont
  {Das~Sarma}}]{liu_andreev_2017}%
  \BibitemOpen
  \bibfield  {author} {\bibinfo {author} {\bibfnamefont {C.-X.}\ \bibnamefont
  {Liu}}, \bibinfo {author} {\bibfnamefont {J.~D.}\ \bibnamefont {Sau}},
  \bibinfo {author} {\bibfnamefont {T.~D.}\ \bibnamefont {Stanescu}},\ and\
  \bibinfo {author} {\bibfnamefont {S.}~\bibnamefont {Das~Sarma}},\ }\bibfield
  {title} {\bibinfo {title} {Andreev bound states versus {Majorana} bound
  states in quantum dot-nanowire-superconductor hybrid structures: {Trivial}
  versus topological zero-bias conductance peaks},\ }\href
  {https://doi.org/10.1103/PhysRevB.96.075161} {\bibfield  {journal} {\bibinfo
  {journal} {Phys. Rev. B}\ }\textbf {\bibinfo {volume} {96}},\ \bibinfo
  {pages} {075161} (\bibinfo {year} {2017})},\ \bibinfo {note} {publisher:
  American Physical Society}\BibitemShut {NoStop}%
\bibitem [{\citenamefont {Liu}\ \emph {et~al.}(2023)\citenamefont {Liu},
  \citenamefont {Pan}, \citenamefont {Setiawan}, \citenamefont {Wimmer},\ and\
  \citenamefont {Sau}}]{liu_fusion_2023}%
  \BibitemOpen
  \bibfield  {author} {\bibinfo {author} {\bibfnamefont {C.-X.}\ \bibnamefont
  {Liu}}, \bibinfo {author} {\bibfnamefont {H.}~\bibnamefont {Pan}}, \bibinfo
  {author} {\bibfnamefont {F.}~\bibnamefont {Setiawan}}, \bibinfo {author}
  {\bibfnamefont {M.}~\bibnamefont {Wimmer}},\ and\ \bibinfo {author}
  {\bibfnamefont {J.~D.}\ \bibnamefont {Sau}},\ }\bibfield  {title} {\bibinfo
  {title} {Fusion protocol for {Majorana} modes in coupled quantum dots},\
  }\href {https://doi.org/10.1103/PhysRevB.108.085437} {\bibfield  {journal}
  {\bibinfo  {journal} {Physical Review B}\ }\textbf {\bibinfo {volume}
  {108}},\ \bibinfo {pages} {085437} (\bibinfo {year} {2023})},\ \bibinfo
  {note} {publisher: American Physical Society}\BibitemShut {NoStop}%
\bibitem [{\citenamefont {Luethi}\ \emph {et~al.}(2023)\citenamefont {Luethi},
  \citenamefont {Legg}, \citenamefont {Laubscher}, \citenamefont {Loss},\ and\
  \citenamefont {Klinovaja}}]{luethi_majorana_2023}%
  \BibitemOpen
  \bibfield  {author} {\bibinfo {author} {\bibfnamefont {M.}~\bibnamefont
  {Luethi}}, \bibinfo {author} {\bibfnamefont {H.~F.}\ \bibnamefont {Legg}},
  \bibinfo {author} {\bibfnamefont {K.}~\bibnamefont {Laubscher}}, \bibinfo
  {author} {\bibfnamefont {D.}~\bibnamefont {Loss}},\ and\ \bibinfo {author}
  {\bibfnamefont {J.}~\bibnamefont {Klinovaja}},\ }\bibfield  {title} {\bibinfo
  {title} {Majorana bound states in germanium {Josephson} junctions via phase
  control},\ }\href {https://doi.org/10.1103/PhysRevB.108.195406} {\bibfield
  {journal} {\bibinfo  {journal} {Physical Review B}\ }\textbf {\bibinfo
  {volume} {108}},\ \bibinfo {pages} {195406} (\bibinfo {year} {2023})},\
  \bibinfo {note} {publisher: American Physical Society}\BibitemShut {NoStop}%
\bibitem [{\citenamefont {Luethi}\ \emph {et~al.}(2024)\citenamefont {Luethi},
  \citenamefont {Legg}, \citenamefont {Loss},\ and\ \citenamefont
  {Klinovaja}}]{luethi_perfect_2024}%
  \BibitemOpen
  \bibfield  {author} {\bibinfo {author} {\bibfnamefont {M.}~\bibnamefont
  {Luethi}}, \bibinfo {author} {\bibfnamefont {H.~F.}\ \bibnamefont {Legg}},
  \bibinfo {author} {\bibfnamefont {D.}~\bibnamefont {Loss}},\ and\ \bibinfo
  {author} {\bibfnamefont {J.}~\bibnamefont {Klinovaja}},\ }\href
  {https://doi.org/10.48550/arXiv.2408.03071} {\bibinfo {title} {From perfect
  to imperfect poor man's {Majoranas} in minimal {Kitaev} chains}} (\bibinfo
  {year} {2024}),\ \bibinfo {note} {arXiv:2408.03071 [cond-mat]}\BibitemShut
  {NoStop}%
\bibitem [{\citenamefont {Bauer}\ \emph {et~al.}(2019)\citenamefont {Bauer},
  \citenamefont {Pereg-Barnea}, \citenamefont {Karzig}, \citenamefont {Rieder},
  \citenamefont {Refael}, \citenamefont {Berg},\ and\ \citenamefont
  {Oreg}}]{bauer_topologically_2019}%
  \BibitemOpen
  \bibfield  {author} {\bibinfo {author} {\bibfnamefont {B.}~\bibnamefont
  {Bauer}}, \bibinfo {author} {\bibfnamefont {T.}~\bibnamefont {Pereg-Barnea}},
  \bibinfo {author} {\bibfnamefont {T.}~\bibnamefont {Karzig}}, \bibinfo
  {author} {\bibfnamefont {M.-T.}\ \bibnamefont {Rieder}}, \bibinfo {author}
  {\bibfnamefont {G.}~\bibnamefont {Refael}}, \bibinfo {author} {\bibfnamefont
  {E.}~\bibnamefont {Berg}},\ and\ \bibinfo {author} {\bibfnamefont
  {Y.}~\bibnamefont {Oreg}},\ }\bibfield  {title} {\bibinfo {title}
  {Topologically protected braiding in a single wire using {Floquet} {Majorana}
  modes},\ }\href {https://doi.org/10.1103/PhysRevB.100.041102} {\bibfield
  {journal} {\bibinfo  {journal} {Physical Review B}\ }\textbf {\bibinfo
  {volume} {100}},\ \bibinfo {pages} {041102} (\bibinfo {year} {2019})},\
  \bibinfo {note} {publisher: American Physical Society}\BibitemShut {NoStop}%
\bibitem [{\citenamefont {Martin}\ and\ \citenamefont
  {Agarwal}(2020)}]{martin_double_2020}%
  \BibitemOpen
  \bibfield  {author} {\bibinfo {author} {\bibfnamefont {I.}~\bibnamefont
  {Martin}}\ and\ \bibinfo {author} {\bibfnamefont {K.}~\bibnamefont
  {Agarwal}},\ }\bibfield  {title} {\bibinfo {title} {Double {Braiding}
  {Majoranas} for {Quantum} {Computing} and {Hamiltonian} {Engineering}},\
  }\href {https://doi.org/10.1103/PRXQuantum.1.020324} {\bibfield  {journal}
  {\bibinfo  {journal} {PRX Quantum}\ }\textbf {\bibinfo {volume} {1}},\
  \bibinfo {pages} {020324} (\bibinfo {year} {2020})},\ \bibinfo {note}
  {publisher: American Physical Society}\BibitemShut {NoStop}%
\bibitem [{\citenamefont {Min}\ \emph {et~al.}(2022)\citenamefont {Min},
  \citenamefont {Fajardo}, \citenamefont {Pereg-Barnea},\ and\ \citenamefont
  {Agarwal}}]{min_dynamical_2022}%
  \BibitemOpen
  \bibfield  {author} {\bibinfo {author} {\bibfnamefont {B.}~\bibnamefont
  {Min}}, \bibinfo {author} {\bibfnamefont {B.}~\bibnamefont {Fajardo}},
  \bibinfo {author} {\bibfnamefont {T.}~\bibnamefont {Pereg-Barnea}},\ and\
  \bibinfo {author} {\bibfnamefont {K.}~\bibnamefont {Agarwal}},\ }\bibfield
  {title} {\bibinfo {title} {Dynamical approach to improving {Majorana} qubits
  and distinguishing them from trivial bound states},\ }\href
  {https://doi.org/10.1103/PhysRevB.105.155412} {\bibfield  {journal} {\bibinfo
   {journal} {Physical Review B}\ }\textbf {\bibinfo {volume} {105}},\ \bibinfo
  {pages} {155412} (\bibinfo {year} {2022})},\ \bibinfo {note} {publisher:
  American Physical Society}\BibitemShut {NoStop}%
\bibitem [{\citenamefont {Lai}\ and\ \citenamefont
  {Zhang}(2020)}]{lai_decoherence_2020}%
  \BibitemOpen
  \bibfield  {author} {\bibinfo {author} {\bibfnamefont {H.-L.}\ \bibnamefont
  {Lai}}\ and\ \bibinfo {author} {\bibfnamefont {W.-M.}\ \bibnamefont
  {Zhang}},\ }\bibfield  {title} {\bibinfo {title} {Decoherence dynamics of
  {Majorana} qubits under braiding operations},\ }\href
  {https://doi.org/10.1103/PhysRevB.101.195428} {\bibfield  {journal} {\bibinfo
   {journal} {Physical Review B}\ }\textbf {\bibinfo {volume} {101}},\ \bibinfo
  {pages} {195428} (\bibinfo {year} {2020})},\ \bibinfo {note} {publisher:
  American Physical Society}\BibitemShut {NoStop}%
\bibitem [{\citenamefont {Boross}\ and\ \citenamefont
  {Pályi}(2022)}]{boross_dephasing_2022}%
  \BibitemOpen
  \bibfield  {author} {\bibinfo {author} {\bibfnamefont {P.}~\bibnamefont
  {Boross}}\ and\ \bibinfo {author} {\bibfnamefont {A.}~\bibnamefont
  {Pályi}},\ }\bibfield  {title} {\bibinfo {title} {Dephasing of {Majorana}
  qubits due to quasistatic disorder},\ }\href
  {https://doi.org/10.1103/PhysRevB.105.035413} {\bibfield  {journal} {\bibinfo
   {journal} {Physical Review B}\ }\textbf {\bibinfo {volume} {105}},\ \bibinfo
  {pages} {035413} (\bibinfo {year} {2022})},\ \bibinfo {note} {publisher:
  American Physical Society}\BibitemShut {NoStop}%
\bibitem [{\citenamefont {Scheurer}\ and\ \citenamefont
  {Shnirman}(2013)}]{scheurer_nonadiabatic_2013}%
  \BibitemOpen
  \bibfield  {author} {\bibinfo {author} {\bibfnamefont {M.~S.}\ \bibnamefont
  {Scheurer}}\ and\ \bibinfo {author} {\bibfnamefont {A.}~\bibnamefont
  {Shnirman}},\ }\bibfield  {title} {\bibinfo {title} {Nonadiabatic processes
  in {Majorana} qubit systems},\ }\href
  {https://doi.org/10.1103/PhysRevB.88.064515} {\bibfield  {journal} {\bibinfo
  {journal} {Physical Review B}\ }\textbf {\bibinfo {volume} {88}},\ \bibinfo
  {pages} {064515} (\bibinfo {year} {2013})},\ \bibinfo {note} {publisher:
  American Physical Society}\BibitemShut {NoStop}%
\bibitem [{\citenamefont {Karzig}\ \emph {et~al.}(2013)\citenamefont {Karzig},
  \citenamefont {Refael},\ and\ \citenamefont {von
  Oppen}}]{karzig_boosting_2013}%
  \BibitemOpen
  \bibfield  {author} {\bibinfo {author} {\bibfnamefont {T.}~\bibnamefont
  {Karzig}}, \bibinfo {author} {\bibfnamefont {G.}~\bibnamefont {Refael}},\
  and\ \bibinfo {author} {\bibfnamefont {F.}~\bibnamefont {von Oppen}},\
  }\bibfield  {title} {\bibinfo {title} {Boosting {Majorana} {Zero} {Modes}},\
  }\href {https://doi.org/10.1103/PhysRevX.3.041017} {\bibfield  {journal}
  {\bibinfo  {journal} {Physical Review X}\ }\textbf {\bibinfo {volume} {3}},\
  \bibinfo {pages} {041017} (\bibinfo {year} {2013})},\ \bibinfo {note}
  {publisher: American Physical Society}\BibitemShut {NoStop}%
\bibitem [{\citenamefont {Karzig}\ \emph
  {et~al.}(2015{\natexlab{a}})\citenamefont {Karzig}, \citenamefont {Rahmani},
  \citenamefont {von Oppen},\ and\ \citenamefont
  {Refael}}]{karzig_optimal_2015}%
  \BibitemOpen
  \bibfield  {author} {\bibinfo {author} {\bibfnamefont {T.}~\bibnamefont
  {Karzig}}, \bibinfo {author} {\bibfnamefont {A.}~\bibnamefont {Rahmani}},
  \bibinfo {author} {\bibfnamefont {F.}~\bibnamefont {von Oppen}},\ and\
  \bibinfo {author} {\bibfnamefont {G.}~\bibnamefont {Refael}},\ }\bibfield
  {title} {\bibinfo {title} {Optimal control of {Majorana} zero modes},\ }\href
  {https://doi.org/10.1103/PhysRevB.91.201404} {\bibfield  {journal} {\bibinfo
  {journal} {Physical Review B}\ }\textbf {\bibinfo {volume} {91}},\ \bibinfo
  {pages} {201404} (\bibinfo {year} {2015}{\natexlab{a}})},\ \bibinfo {note}
  {publisher: American Physical Society}\BibitemShut {NoStop}%
\bibitem [{\citenamefont {Bauer}\ \emph {et~al.}(2018)\citenamefont {Bauer},
  \citenamefont {Karzig}, \citenamefont {Mishmash}, \citenamefont {Antipov},\
  and\ \citenamefont {Alicea}}]{bauer_dynamics_2018}%
  \BibitemOpen
  \bibfield  {author} {\bibinfo {author} {\bibfnamefont {B.}~\bibnamefont
  {Bauer}}, \bibinfo {author} {\bibfnamefont {T.}~\bibnamefont {Karzig}},
  \bibinfo {author} {\bibfnamefont {R.}~\bibnamefont {Mishmash}}, \bibinfo
  {author} {\bibfnamefont {A.}~\bibnamefont {Antipov}},\ and\ \bibinfo {author}
  {\bibfnamefont {J.}~\bibnamefont {Alicea}},\ }\bibfield  {title} {\bibinfo
  {title} {Dynamics of majorana-based qubits operated with an array of tunable
  gates},\ }\href {https://doi.org/10.21468/SciPostPhys.5.1.004} {\bibfield
  {journal} {\bibinfo  {journal} {SciPost Physics}\ }\textbf {\bibinfo {volume}
  {5}},\ \bibinfo {pages} {004} (\bibinfo {year} {2018})}\BibitemShut {NoStop}%
\bibitem [{\citenamefont {Conlon}\ \emph {et~al.}(2019)\citenamefont {Conlon},
  \citenamefont {Pellegrino}, \citenamefont {Slingerland}, \citenamefont
  {Dooley},\ and\ \citenamefont {Kells}}]{conlon_error_2019}%
  \BibitemOpen
  \bibfield  {author} {\bibinfo {author} {\bibfnamefont {A.}~\bibnamefont
  {Conlon}}, \bibinfo {author} {\bibfnamefont {D.}~\bibnamefont {Pellegrino}},
  \bibinfo {author} {\bibfnamefont {J.~K.}\ \bibnamefont {Slingerland}},
  \bibinfo {author} {\bibfnamefont {S.}~\bibnamefont {Dooley}},\ and\ \bibinfo
  {author} {\bibfnamefont {G.}~\bibnamefont {Kells}},\ }\bibfield  {title}
  {\bibinfo {title} {Error generation and propagation in {Majorana}-based
  topological qubits},\ }\href {https://doi.org/10.1103/PhysRevB.100.134307}
  {\bibfield  {journal} {\bibinfo  {journal} {Physical Review B}\ }\textbf
  {\bibinfo {volume} {100}},\ \bibinfo {pages} {134307} (\bibinfo {year}
  {2019})},\ \bibinfo {note} {publisher: American Physical Society}\BibitemShut
  {NoStop}%
\bibitem [{\citenamefont {Coopmans}\ \emph {et~al.}(2021)\citenamefont
  {Coopmans}, \citenamefont {Luo}, \citenamefont {Kells}, \citenamefont
  {Clark},\ and\ \citenamefont {Carrasquilla}}]{coopmans_protocol_2021}%
  \BibitemOpen
  \bibfield  {author} {\bibinfo {author} {\bibfnamefont {L.}~\bibnamefont
  {Coopmans}}, \bibinfo {author} {\bibfnamefont {D.}~\bibnamefont {Luo}},
  \bibinfo {author} {\bibfnamefont {G.}~\bibnamefont {Kells}}, \bibinfo
  {author} {\bibfnamefont {B.~K.}\ \bibnamefont {Clark}},\ and\ \bibinfo
  {author} {\bibfnamefont {J.}~\bibnamefont {Carrasquilla}},\ }\bibfield
  {title} {\bibinfo {title} {Protocol {Discovery} for the {Quantum} {Control}
  of {Majoranas} by {Differentiable} {Programming} and {Natural} {Evolution}
  {Strategies}},\ }\href {https://doi.org/10.1103/PRXQuantum.2.020332}
  {\bibfield  {journal} {\bibinfo  {journal} {PRX Quantum}\ }\textbf {\bibinfo
  {volume} {2}},\ \bibinfo {pages} {020332} (\bibinfo {year} {2021})},\
  \bibinfo {note} {publisher: American Physical Society}\BibitemShut {NoStop}%
\bibitem [{\citenamefont {Xu}\ and\ \citenamefont
  {Li}(2022)}]{xu_transport_2022}%
  \BibitemOpen
  \bibfield  {author} {\bibinfo {author} {\bibfnamefont {L.}~\bibnamefont
  {Xu}}\ and\ \bibinfo {author} {\bibfnamefont {X.-Q.}\ \bibnamefont {Li}},\
  }\bibfield  {title} {\bibinfo {title} {Transport probe of the nonadiabatic
  transition caused by moving {Majorana} zero modes},\ }\href
  {https://doi.org/10.1103/PhysRevB.105.245410} {\bibfield  {journal} {\bibinfo
   {journal} {Physical Review B}\ }\textbf {\bibinfo {volume} {105}},\ \bibinfo
  {pages} {245410} (\bibinfo {year} {2022})},\ \bibinfo {note} {publisher:
  American Physical Society}\BibitemShut {NoStop}%
\bibitem [{\citenamefont {Nag}\ and\ \citenamefont
  {Sau}(2019)}]{nag_diabatic_2019}%
  \BibitemOpen
  \bibfield  {author} {\bibinfo {author} {\bibfnamefont {A.}~\bibnamefont
  {Nag}}\ and\ \bibinfo {author} {\bibfnamefont {J.~D.}\ \bibnamefont {Sau}},\
  }\bibfield  {title} {\bibinfo {title} {Diabatic errors in {Majorana} braiding
  with bosonic bath},\ }\href {https://doi.org/10.1103/PhysRevB.100.014511}
  {\bibfield  {journal} {\bibinfo  {journal} {Physical Review B}\ }\textbf
  {\bibinfo {volume} {100}},\ \bibinfo {pages} {014511} (\bibinfo {year}
  {2019})},\ \bibinfo {note} {publisher: American Physical Society}\BibitemShut
  {NoStop}%
\bibitem [{\citenamefont {Mascot}\ \emph {et~al.}(2023)\citenamefont {Mascot},
  \citenamefont {Hodge}, \citenamefont {Crawford}, \citenamefont {Bedow},
  \citenamefont {Morr},\ and\ \citenamefont {Rachel}}]{mascot_many-body_2023}%
  \BibitemOpen
  \bibfield  {author} {\bibinfo {author} {\bibfnamefont {E.}~\bibnamefont
  {Mascot}}, \bibinfo {author} {\bibfnamefont {T.}~\bibnamefont {Hodge}},
  \bibinfo {author} {\bibfnamefont {D.}~\bibnamefont {Crawford}}, \bibinfo
  {author} {\bibfnamefont {J.}~\bibnamefont {Bedow}}, \bibinfo {author}
  {\bibfnamefont {D.~K.}\ \bibnamefont {Morr}},\ and\ \bibinfo {author}
  {\bibfnamefont {S.}~\bibnamefont {Rachel}},\ }\bibfield  {title} {\bibinfo
  {title} {Many-{Body} {Majorana} {Braiding} without an {Exponential} {Hilbert}
  {Space}},\ }\href {https://doi.org/10.1103/PhysRevLett.131.176601} {\bibfield
   {journal} {\bibinfo  {journal} {Physical Review Letters}\ }\textbf {\bibinfo
  {volume} {131}},\ \bibinfo {pages} {176601} (\bibinfo {year} {2023})},\
  \bibinfo {note} {publisher: American Physical Society}\BibitemShut {NoStop}%
\bibitem [{\citenamefont {Cheng}\ \emph {et~al.}(2011)\citenamefont {Cheng},
  \citenamefont {Galitski},\ and\ \citenamefont
  {Das~Sarma}}]{cheng_nonadiabatic_2011}%
  \BibitemOpen
  \bibfield  {author} {\bibinfo {author} {\bibfnamefont {M.}~\bibnamefont
  {Cheng}}, \bibinfo {author} {\bibfnamefont {V.}~\bibnamefont {Galitski}},\
  and\ \bibinfo {author} {\bibfnamefont {S.}~\bibnamefont {Das~Sarma}},\
  }\bibfield  {title} {\bibinfo {title} {Nonadiabatic effects in the braiding
  of non-{Abelian} anyons in topological superconductors},\ }\href
  {https://doi.org/10.1103/PhysRevB.84.104529} {\bibfield  {journal} {\bibinfo
  {journal} {Physical Review B}\ }\textbf {\bibinfo {volume} {84}},\ \bibinfo
  {pages} {104529} (\bibinfo {year} {2011})},\ \bibinfo {note} {publisher:
  American Physical Society}\BibitemShut {NoStop}%
\bibitem [{\citenamefont {Karzig}\ \emph
  {et~al.}(2015{\natexlab{b}})\citenamefont {Karzig}, \citenamefont {Pientka},
  \citenamefont {Refael},\ and\ \citenamefont {von
  Oppen}}]{karzig_shortcuts_2015}%
  \BibitemOpen
  \bibfield  {author} {\bibinfo {author} {\bibfnamefont {T.}~\bibnamefont
  {Karzig}}, \bibinfo {author} {\bibfnamefont {F.}~\bibnamefont {Pientka}},
  \bibinfo {author} {\bibfnamefont {G.}~\bibnamefont {Refael}},\ and\ \bibinfo
  {author} {\bibfnamefont {F.}~\bibnamefont {von Oppen}},\ }\bibfield  {title}
  {\bibinfo {title} {Shortcuts to non-{Abelian} braiding},\ }\href
  {https://doi.org/10.1103/PhysRevB.91.201102} {\bibfield  {journal} {\bibinfo
  {journal} {Physical Review B}\ }\textbf {\bibinfo {volume} {91}},\ \bibinfo
  {pages} {201102} (\bibinfo {year} {2015}{\natexlab{b}})},\ \bibinfo {note}
  {publisher: American Physical Society}\BibitemShut {NoStop}%
\bibitem [{\citenamefont {Knapp}\ \emph {et~al.}(2016)\citenamefont {Knapp},
  \citenamefont {Zaletel}, \citenamefont {Liu}, \citenamefont {Cheng},
  \citenamefont {Bonderson},\ and\ \citenamefont {Nayak}}]{knapp_nature_2016}%
  \BibitemOpen
  \bibfield  {author} {\bibinfo {author} {\bibfnamefont {C.}~\bibnamefont
  {Knapp}}, \bibinfo {author} {\bibfnamefont {M.}~\bibnamefont {Zaletel}},
  \bibinfo {author} {\bibfnamefont {D.~E.}\ \bibnamefont {Liu}}, \bibinfo
  {author} {\bibfnamefont {M.}~\bibnamefont {Cheng}}, \bibinfo {author}
  {\bibfnamefont {P.}~\bibnamefont {Bonderson}},\ and\ \bibinfo {author}
  {\bibfnamefont {C.}~\bibnamefont {Nayak}},\ }\bibfield  {title} {\bibinfo
  {title} {The {Nature} and {Correction} of {Diabatic} {Errors} in {Anyon}
  {Braiding}},\ }\href {https://doi.org/10.1103/PhysRevX.6.041003} {\bibfield
  {journal} {\bibinfo  {journal} {Physical Review X}\ }\textbf {\bibinfo
  {volume} {6}},\ \bibinfo {pages} {041003} (\bibinfo {year} {2016})},\
  \bibinfo {note} {publisher: American Physical Society}\BibitemShut {NoStop}%
\bibitem [{\citenamefont {Hell}\ \emph {et~al.}(2016)\citenamefont {Hell},
  \citenamefont {Danon}, \citenamefont {Flensberg},\ and\ \citenamefont
  {Leijnse}}]{hell_time_2016}%
  \BibitemOpen
  \bibfield  {author} {\bibinfo {author} {\bibfnamefont {M.}~\bibnamefont
  {Hell}}, \bibinfo {author} {\bibfnamefont {J.}~\bibnamefont {Danon}},
  \bibinfo {author} {\bibfnamefont {K.}~\bibnamefont {Flensberg}},\ and\
  \bibinfo {author} {\bibfnamefont {M.}~\bibnamefont {Leijnse}},\ }\bibfield
  {title} {\bibinfo {title} {Time scales for {Majorana} manipulation using
  {Coulomb} blockade in gate-controlled superconducting nanowires},\ }\href
  {https://doi.org/10.1103/PhysRevB.94.035424} {\bibfield  {journal} {\bibinfo
  {journal} {Physical Review B}\ }\textbf {\bibinfo {volume} {94}},\ \bibinfo
  {pages} {035424} (\bibinfo {year} {2016})},\ \bibinfo {note} {publisher:
  American Physical Society}\BibitemShut {NoStop}%
\bibitem [{\citenamefont {Rahmani}\ \emph {et~al.}(2017)\citenamefont
  {Rahmani}, \citenamefont {Seradjeh},\ and\ \citenamefont
  {Franz}}]{rahmani_optimal_2017}%
  \BibitemOpen
  \bibfield  {author} {\bibinfo {author} {\bibfnamefont {A.}~\bibnamefont
  {Rahmani}}, \bibinfo {author} {\bibfnamefont {B.}~\bibnamefont {Seradjeh}},\
  and\ \bibinfo {author} {\bibfnamefont {M.}~\bibnamefont {Franz}},\ }\bibfield
   {title} {\bibinfo {title} {Optimal diabatic dynamics of {Majorana}-based
  quantum gates},\ }\href {https://doi.org/10.1103/PhysRevB.96.075158}
  {\bibfield  {journal} {\bibinfo  {journal} {Physical Review B}\ }\textbf
  {\bibinfo {volume} {96}},\ \bibinfo {pages} {075158} (\bibinfo {year}
  {2017})},\ \bibinfo {note} {publisher: American Physical Society}\BibitemShut
  {NoStop}%
\bibitem [{\citenamefont {Sekania}\ \emph {et~al.}(2017)\citenamefont
  {Sekania}, \citenamefont {Plugge}, \citenamefont {Greiter}, \citenamefont
  {Thomale},\ and\ \citenamefont {Schmitteckert}}]{sekania_braiding_2017}%
  \BibitemOpen
  \bibfield  {author} {\bibinfo {author} {\bibfnamefont {M.}~\bibnamefont
  {Sekania}}, \bibinfo {author} {\bibfnamefont {S.}~\bibnamefont {Plugge}},
  \bibinfo {author} {\bibfnamefont {M.}~\bibnamefont {Greiter}}, \bibinfo
  {author} {\bibfnamefont {R.}~\bibnamefont {Thomale}},\ and\ \bibinfo {author}
  {\bibfnamefont {P.}~\bibnamefont {Schmitteckert}},\ }\bibfield  {title}
  {\bibinfo {title} {Braiding errors in interacting {Majorana} quantum wires},\
  }\href {https://doi.org/10.1103/PhysRevB.96.094307} {\bibfield  {journal}
  {\bibinfo  {journal} {Physical Review B}\ }\textbf {\bibinfo {volume} {96}},\
  \bibinfo {pages} {094307} (\bibinfo {year} {2017})},\ \bibinfo {note}
  {publisher: American Physical Society}\BibitemShut {NoStop}%
\bibitem [{\citenamefont {Zhang}\ \emph {et~al.}(2019)\citenamefont {Zhang},
  \citenamefont {Mei}, \citenamefont {Meng}, \citenamefont {Liang},\ and\
  \citenamefont {Yang}}]{zhang_effects_2019}%
  \BibitemOpen
  \bibfield  {author} {\bibinfo {author} {\bibfnamefont {Z.-T.}\ \bibnamefont
  {Zhang}}, \bibinfo {author} {\bibfnamefont {F.}~\bibnamefont {Mei}}, \bibinfo
  {author} {\bibfnamefont {X.-G.}\ \bibnamefont {Meng}}, \bibinfo {author}
  {\bibfnamefont {B.-L.}\ \bibnamefont {Liang}},\ and\ \bibinfo {author}
  {\bibfnamefont {Z.-S.}\ \bibnamefont {Yang}},\ }\bibfield  {title} {\bibinfo
  {title} {Effects of decoherence on diabatic errors in {Majorana} braiding},\
  }\href {https://doi.org/10.1103/PhysRevA.100.012324} {\bibfield  {journal}
  {\bibinfo  {journal} {Physical Review A}\ }\textbf {\bibinfo {volume}
  {100}},\ \bibinfo {pages} {012324} (\bibinfo {year} {2019})},\ \bibinfo
  {note} {publisher: American Physical Society}\BibitemShut {NoStop}%
\bibitem [{\citenamefont {Truong}\ \emph {et~al.}(2023)\citenamefont {Truong},
  \citenamefont {Agarwal},\ and\ \citenamefont
  {Pereg-Barnea}}]{truong_optimizing_2023}%
  \BibitemOpen
  \bibfield  {author} {\bibinfo {author} {\bibfnamefont {B.~P.}\ \bibnamefont
  {Truong}}, \bibinfo {author} {\bibfnamefont {K.}~\bibnamefont {Agarwal}},\
  and\ \bibinfo {author} {\bibfnamefont {T.}~\bibnamefont {Pereg-Barnea}},\
  }\bibfield  {title} {\bibinfo {title} {Optimizing the transport of {Majorana}
  zero modes in one-dimensional topological superconductors},\ }\href
  {https://doi.org/10.1103/PhysRevB.107.104516} {\bibfield  {journal} {\bibinfo
   {journal} {Physical Review B}\ }\textbf {\bibinfo {volume} {107}},\ \bibinfo
  {pages} {104516} (\bibinfo {year} {2023})},\ \bibinfo {note} {publisher:
  American Physical Society}\BibitemShut {NoStop}%
\bibitem [{\citenamefont {Wang}\ \emph {et~al.}(2024)\citenamefont {Wang},
  \citenamefont {Bai}, \citenamefont {Xu}, \citenamefont {Feng},\ and\
  \citenamefont {Li}}]{wang_transport_2024}%
  \BibitemOpen
  \bibfield  {author} {\bibinfo {author} {\bibfnamefont {Q.}~\bibnamefont
  {Wang}}, \bibinfo {author} {\bibfnamefont {J.}~\bibnamefont {Bai}}, \bibinfo
  {author} {\bibfnamefont {L.}~\bibnamefont {Xu}}, \bibinfo {author}
  {\bibfnamefont {W.}~\bibnamefont {Feng}},\ and\ \bibinfo {author}
  {\bibfnamefont {X.-Q.}\ \bibnamefont {Li}},\ }\bibfield  {title} {\bibinfo
  {title} {Transport and fusion of {Majorana} zero modes in the presence of
  nonadiabatic transitions},\ }\href
  {https://doi.org/10.1103/PhysRevB.110.115402} {\bibfield  {journal} {\bibinfo
   {journal} {Physical Review B}\ }\textbf {\bibinfo {volume} {110}},\ \bibinfo
  {pages} {115402} (\bibinfo {year} {2024})}\BibitemShut {NoStop}%
\bibitem [{\citenamefont {Hegde}\ \emph {et~al.}(2015)\citenamefont {Hegde},
  \citenamefont {Shivamoggi}, \citenamefont {Vishveshwara},\ and\ \citenamefont
  {Sen}}]{hegde_quench_2015}%
  \BibitemOpen
  \bibfield  {author} {\bibinfo {author} {\bibfnamefont {S.}~\bibnamefont
  {Hegde}}, \bibinfo {author} {\bibfnamefont {V.}~\bibnamefont {Shivamoggi}},
  \bibinfo {author} {\bibfnamefont {S.}~\bibnamefont {Vishveshwara}},\ and\
  \bibinfo {author} {\bibfnamefont {D.}~\bibnamefont {Sen}},\ }\bibfield
  {title} {\bibinfo {title} {Quench dynamics and parity blocking in {Majorana}
  wires},\ }\href@noop {} {\bibfield  {journal} {\bibinfo  {journal} {New
  Journal of Physics}\ }\textbf {\bibinfo {volume} {17}},\ \bibinfo {pages}
  {053036} (\bibinfo {year} {2015})},\ \bibinfo {note} {publisher: IOP
  Publishing}\BibitemShut {NoStop}%
\bibitem [{\citenamefont {Xu}\ \emph {et~al.}(2023)\citenamefont {Xu},
  \citenamefont {Bai}, \citenamefont {Feng},\ and\ \citenamefont
  {Li}}]{xu_dynamics_2023}%
  \BibitemOpen
  \bibfield  {author} {\bibinfo {author} {\bibfnamefont {L.}~\bibnamefont
  {Xu}}, \bibinfo {author} {\bibfnamefont {J.}~\bibnamefont {Bai}}, \bibinfo
  {author} {\bibfnamefont {W.}~\bibnamefont {Feng}},\ and\ \bibinfo {author}
  {\bibfnamefont {X.-Q.}\ \bibnamefont {Li}},\ }\bibfield  {title} {\bibinfo
  {title} {Dynamics simulation of braiding two {Majorana} zero modes via a
  quantum dot},\ }\href {https://doi.org/10.1103/PhysRevB.108.115411}
  {\bibfield  {journal} {\bibinfo  {journal} {Phys. Rev. B}\ }\textbf {\bibinfo
  {volume} {108}},\ \bibinfo {pages} {115411} (\bibinfo {year} {2023})},\
  \bibinfo {note} {publisher: American Physical Society}\BibitemShut {NoStop}%
\end{thebibliography}%

\end{document}